\documentclass{elsart}
\usepackage[dvips]{epsfig}
\usepackage{amsmath}

\newlength{\wth}
\begin{document}

\begin{frontmatter}

\title{Sampling using a `bank' of clues}

\author[adben]{Benjamin C Allanach},
\corauth[coboth]{Corresponding author.}
\ead{benjamin.allanach@cern.ch}
\author[adchr]{Christopher G Lester\corauthref{coboth}} 
\ead{christopher.lester@cern.ch}
\address[adben]{DAMTP, CMS, Wilberforce Road, Cambridge CB3 0WA, UK}
\address[adchr]{Cavendish Laboratory, J.J. Thomson Avenue, Cambridge CB3 0HE, UK}

\begin{abstract}
An easy-to-implement form of the Metropolis Algorithm is described
which, unlike most standard techniques, is well suited to sampling
from multi-modal distributions on spaces with moderate numbers of
dimensions (order ten) in environments typical of investigations into
current constraints on Beyond-the-Standard-Model physics. The sampling
technique makes use of pre-existing information (which can safely be
of low or uncertain quality) relating to the distribution from which
it is desired to sample.  This information should come in the form of
a ``bank'' or ``cache'' of parameter space points of which {\em at
least some} may be expected to be near regions of interest in the
desired distribution.  In practical circumstances such ``banks of
clues'' are easy to assemble from earlier work, aborted runs,
discarded burn-in samples from failed sampling attempts, or from prior
scouting investigations. The technique equilibrates between
disconnected parts of the distribution without user input. The
algorithm is not lead astray by ``bad'' clues, but there is no free
lunch: performance gains will only be seen where clues are helpful.
\end{abstract}


\end{frontmatter}

\newcommand{\fourgraphs}[4]{%
 \unitlength=1.1in
 \begin{picture}(5.8,4.4)(0.3,0.3)
\put(0,2.4){\put(0.5,0){\epsfig{file=#1, width=0.698 \wth}}
 \put(2.9,0){\epsfig{file=#2, width=0.698 \wth}}
 \put(0.5,2.2){(a)}
 \put(3,2.2){(b)}}
\put(0,0){\put(0.5,0){\epsfig{file=#3, width=0.698 \wth}}
 \put(2.9,0){\epsfig{file=#4, width=0.698 \wth}}
 \put(0.5,2.2){(c)}
 \put(3,2.2){(d)}}
 \end{picture}
}
\newcommand{\sixgraphs}[6]{%
\unitlength=0.86in
\begin{picture}(5,7.5)
\put(0,5){\epsfig{file=#1, width=2.15in}}
\put(2.5,5){\epsfig{file=#2, width=2.15in}}
\put(0,0){\epsfig{file=#5, width=2.15in}}
\put(2.5,0){\epsfig{file=#6, width=2.15in}}
\put(0,2.5){\epsfig{file=#3, width=2.15in}}
\put(2.5,2.5){\epsfig{file=#4, width=2.15in}}
\put(0,4.7){(c)}
\put(0,7.2){(a)}
\put(2.5,7.2){(b)}
\put(0,2.2){(e)}
\put(2.5,2.2){(f)}
\put(2.5,4.7){(d)}
\end{picture}}

\newcommand{\threegraphst}[3]{%
\unitlength=0.75in
\begin{picture}(7.5,3)(0.5,0.25)
\put(0.0,0){\epsfig{file=#1, width=2.05in}}
\put(2.5,0){\epsfig{file=#2, width=2.05in}}
\put(5.0,0){\epsfig{file=#3, width=2.05in}}
\put(1.1,2.5){(a)}
\put(3.6,2.5){(b)}
\put(6.1,2.5){(c)}
\end{picture}}

\newcommand{\ninegraphs}[9]{%
\unitlength=0.75in
\begin{picture}(7.5,9)
\put(0.0,6){\epsfig{file=#1, width=2.05in}}
\put(2.5,6){\epsfig{file=#2, width=2.05in}}
\put(5.0,6){\epsfig{file=#3, width=2.05in}}
\put(0.0,3){\epsfig{file=#4, width=2.05in}}
\put(2.5,3){\epsfig{file=#5, width=2.05in}}
\put(5.0,3){\epsfig{file=#6, width=2.05in}}
\put(0.0,0){\epsfig{file=#7, width=2.05in}}
\put(2.5,0){\epsfig{file=#8, width=2.05in}}
\put(5.0,0){\epsfig{file=#9, width=2.05in}}
\put(1.1,8.5){(a)}
\put(3.6,8.5){(b)}
\put(6.1,8.5){(c)}
\put(1.1,5.5){(d)}
\put(3.6,5.5){(e)}
\put(6.1,5.5){(f)}
\put(1.1,2.5){(g)}
\put(3.6,2.5){(h)}
\put(6.1,2.5){(i)}
\end{picture}}

\newcommand{\fourgraphst}[4]{%
 \unitlength=1.1in
 \begin{picture}(5.8,4)(0.5,0.4)
\put(0,2){\put(-0.04,2.54){\epsfig{file=#1, width=0.698 \wth,angle=270}}
 \put(0.85,0.5){\epsfig{file=#12, width=0.68 \wth}}
 \put(2.66,2.54){\epsfig{file=#2, width=0.698 \wth, angle=270}}
 \put(3.55,0.5){\epsfig{file=#22, width=0.68 \wth}}
 \put(0.5,2.1){(a)}
 \put(3.2,2.1){(b)}
}
\put(0,0){\put(-0.04,2.54){\epsfig{file=#3, width=0.698 \wth,angle=270}}
 \put(0.85,0.5){\epsfig{file=#32, width=0.68 \wth}}
 \put(2.66,2.54){\epsfig{file=#4, width=0.698 \wth, angle=270}}
 \put(3.55,0.5){\epsfig{file=#42, width=0.68 \wth}}
 \put(0.5,2.1){(c)}
 \put(3.2,2.1){(d)}
}
 \end{picture}
}

\newcommand{\fourgraphstt}[4]{%
 \unitlength=1.1in
 \begin{picture}(5.8,4)(0.5,0.4)
\put(0,2){
  \put(0.45,0.15){\epsfig{file=#1, width=0.6 \wth}}
  \put(2.66,2.54){\epsfig{file=#2, width=0.698 \wth, angle=270}}
  \put(3.55,0.5){\epsfig{file=#22, width=0.68 \wth}}
  \put(0.5,2.1){(a)}
  \put(3.2,2.1){(b)}
}
\put(0,0){\put(-0.04,2.54){\epsfig{file=#3, width=0.698 \wth,angle=270}}
 \put(0.85,0.5){\epsfig{file=#32, width=0.68 \wth}}
 \put(2.66,2.54){\epsfig{file=#4, width=0.698 \wth, angle=270}}
 \put(3.55,0.5){\epsfig{file=#42, width=0.68 \wth}}
 \put(0.5,2.1){(c)}
 \put(3.2,2.1){(d)}
}
 \end{picture}
}

\newcommand{\twographst}[2]{%
 \unitlength=1.1in
 \begin{picture}(5.8,2.3)(0.5,0.25)
 \put(0.7,0.3){\epsfig{file=#1, width=0.6 \wth}}
 \put(3.56,0.3){\epsfig{file=#2, width=0.6 \wth}}
 \put(0.5,2.1){(a)}
 \put(3.2,2.1){(b)}
 \end{picture}
}
\newcommand{\twographs}[2]{%
 \unitlength=1.1in
 \begin{picture}(5.8,2.3)(0.5,0.25)
 \put(0.85,0.5){\epsfig{file=#1, width=0.68 \wth}}
 \put(3.56,0.5){\epsfig{file=#2, width=0.68 \wth}}
 \put(0.5,2.1){(a)}
 \put(3.2,2.1){(b)}
 \end{picture}
}

\newcommand{\eightgraphs}[8]{%
 \unitlength=1in
 \begin{picture}(6,6)(0,0)
\put(0,4){\epsfig{file=#1, width=2 in}}
\put(2,4){\epsfig{file=#2, width=2 in}}
\put(4,4){\epsfig{file=#3, width=2 in}}
\put(0,2){\epsfig{file=#4, width=2 in}}
\put(2,2){\epsfig{file=#5, width=2 in}}
\put(4,2){\epsfig{file=#6, width=2 in}}
\put(1,0){\epsfig{file=#7, width=2 in}}
\put(3,0){\epsfig{file=#82, width=2 in}}
\put(3.8,0.6){\epsfig{file=#81, width=1 in}}
\put(0,5.8){(a)}
\put(2,5.8){(b)}
\put(4,5.8){(c)}
\put(0,3.8){(d)}
\put(2,3.8){(e)}
\put(4,3.8){(f)}
\put(1,1.8){(g)}
\put(3,1.8){(h)}
 \end{picture}
}

\newcommand{\circf}{\mathop{\mathrm{circ}}}


\def\mygloss#1#2#3{\glossary{#1 @{\bf #2} & #3}}
\newcommand{\prepareAbbrev}[6]{\newcounter{#5}\newcommand{#1}{\mygloss{#2}{#6}{#4}\ifnum\arabic{#5}=0 {#4 (#3)}\else#3\fi\addtocounter{#5}{1}}}
\prepareAbbrev\MHA{MHA}{MHA}{Metropolis-Hastings Algorithm~\cite{Metropolis,Hastings,MacKay}}{MHAcounter}{MHA}
\prepareAbbrev\MCMC{MCMC}{MCMC}{Markov Chain Monte Carlo}{MCMCcounter}{MCMC}
\prepareAbbrev\MHMC{MHMC}{MHMC}{Metropolis-Hastings Monte Carlo}{MHMCcounter}{MHMC}

\newcommand\PD{{proposal distribution}}

\newcommand\rhatmax{{\hat R_{max}}}

\section{Introduction}

\subsection{Exploring parameter spaces of moderate dimension}

Many scientific disciplines share, from time to time, the need to
explore parameter spaces with moderate-to-large numbers of dimensions
(that is to say tens of parameters, but not thousands of parameters).
It is not uncommon to find that the only algorithms capable of
exploring these spaces within reasonable timescales are \MCMC\
methods.  One such method of particular historical importance is the
\MHA\ which provides a simple way of sampling parts of the space in
proportion to a measure defined on that space.  There is a freedom
available to anyone implementing the \MHA\ for a given problem, and
this is the freedom to choose the so-called \PD\ $Q(x|y)$.

\subsection{Making exploration efficient}

The \MHA\ and \PD s are described in more detail in
section~\ref{sec:gensec2}.  Here we need only note that the issue of
choice regarding the \PD\ is one which is intimately connected to the
efficiency\footnote{In this context a more ``efficient'' algorithm is
one which needs to be run for a smaller number of iterations than a
competitor, before the samples may be regarded as independent samples
from the full distribution.} of an implementation in a particular
problem.  The choice of \PD\ does not affect the quality of the
samples obtained from the \MHA\ once the sampling has run for long
enough.  However, a judicious choice of \PD\ may reduce the time
needed to wait for the samples to become effectively independent and
representative of the bulk of the probability mass by many orders of
magnitude.  Careful choice of a \PD\ can therefore allow more
complicated problems to be attacked, or allow existing problems to be
solved in shorter times.


\subsection{The purpose of this paper}

We describe a new ``bank sampling'' algorithm created specifically to
address the problems associated with sampling from distributions
containing isolated modes (regions of interest).  The method addresses
some of the failings, detailed above, of the standard \MHMC\
techniques which have been used in particle physics up to the present.
The most important claim is that the new algorithm moves between
isolated modes freely, and can thereby establish their relative
weights at no great extra cost to the user.

We do not claim that the method would be regarded as novel by the
statistical community.  In section \ref{sec:pains} we will make clear
that the ``bank sampling'' algorithm is just a special case of the
\MHA.  Algorithms of the same type have previously been described as
``mixture hybrid kernel \MCMC'' or ``mixture strategy \MCMC'' methods
(see for example \cite{Tierney}) and have been around for many years.
The novel component of the algorithm is (1) its application to
particle physics parameter space investigations, which have
historically been slow to follow the statistical literature, and (2)
the suggested nature in which the algorithm aims to make use of data
which is commonly produced during the ``lead in'' to parameter space
scans, but which is usually discarded.

The method does {\em not} make any claim to be able to {\em discover}
isolated regions.\footnote{Bank sampling is not in competition
with techniques such as Annealed Importance Sampling, Simulated
Annealing and their relations (\cite{neal-1998} and references
therein).}  On the contrary, the method is only an improvement on
standard \MHMC\ techniques if something is already known about the
location of the isolated regions -- perhaps revealed by an earlier
standard \MHMC\ investigation.  If one of the modes is left out of the
bank points, the sampler only has a small chance per step of reaching
it from one of the other modes. If the number of modes is very large
(for example in the thousands), then it is likely that another method
such as simulated annealing would provide a more reliable sampler.

\section{The Bank Sampling Algorithm}
\label{sec:gensec2}
The ``bank sampling'' algorithm may be classed as a \MHA\ with
a particular choice of \PD.  We must therefore begin by describing the
\MHA\ and the role of the \PD, before then going on to describe the
particular \PD\ used in bank sampling.

\subsection{The Metropolis-Hastings Algorithm}

The \MHA\ generates a sequence of points $\left\{ {\mathbf x}^{(t)}
\right\}$ in a parameter space on which a distribution
${\mathcal L}({\mathbf x})$ is defined.  Here $t$ labels the position
of the point within the sequence.  We will refer to this sequence of
points as the ``Markov chain''.  The \MHA\ extends a chain of length
$t$ to a chain of length $t+1$ in the following way.

\label{sec:sec:propdist}

A new point ${\mathbf x'}$, not necessarily destined to become
${\mathbf x}^{(t+1)}$, is chosen from a proposal distribution
$Q({\mathbf x};{\mathbf x}^{(t)})$ that is allowed to depend on
${\mathbf x}^{(t)}$ but not on any of $\left\{ {\mathbf x}^{(t-1)},
{\mathbf x}^{(t-2)}, ... \right\}$.  It is this \PD\ which will take on a
particular form in bank sampling.  The ratio
\begin{equation}
\label{eq:rho}
\rho = 
\frac{{\mathcal L}({\bf x'})}{{\mathcal L}{({\bf x}^{(t)} )}} \cdot
\frac{Q({\mathbf x}^{(t)};{\mathbf x'})}{Q({\mathbf x'};{\mathbf x}^{(t)})}
\end{equation} 
is then computed, and used to decide whether to accept or reject the
proposal. If the ratio $\rho$ is greater than or equal to one, the new
point ${\bf x}$ is accepted and appended to the chain. If $\rho$ is
less than one, the new point ${\bf x'}$ is only accepted and appended
to the chain with probability $\rho$.\footnote{For example a random
number $\lambda$ may be drawn uniformly from the unit interval and
${\bf x'}$ appended to the chain only if $\lambda<\rho$.}  If neither
of these tests succeeds in appending ${\bf x'}$ to the chain, then the
old point ${\mathbf x}^{(t)}$ is appended instead.  Whichever of ${\bf
x'}$ or ${\mathbf x}^{(t)}$ makes it on to the end of the chain is
thereafter known as ${\mathbf x}^{(t+1)}$.

As a result of following the above steps, the sampling density of
points in the chain becomes proportional to the density of the target
distribution ${\mathcal L}({\mathbf x})$ as the number of links goes
to infinity.  In the limit of very large times, the result becomes
independent of $Q$ (the \PD) provided that it meets some reasonable
conditions described in \cite{MacKay}.

\subsection{The \PD\ for the Bank Sampler} \label{sec:pains}
What is needed is a \PD\ that combines the best of both local and
global forms - and this is what the \PD\ of the bank sampler in
equation~(\ref{eq:bankpd}) attempts to do.

Given a ``bank'' of $N$ points ${\mathbf y}^{(i)}$, some of which have
been previously identified as lying in or near regions of interest in
the space, perhaps as a result of earlier analyses, one can construct
the following \PD:
\begin{equation}
Q_{bank}(     {\mathbf x};{\mathbf x}^{(t)}       ) =
(1-\lambda) K_0( {\mathbf x};{\mathbf x}^{(t)}       ) + 
\lambda \sum_{i=1}^N {
w_i K_i(     {\mathbf x};{\mathbf y}^{(i)}       )
} \label{eq:bankpd}
\end{equation}
in which $w_i$ are a set of $N$ weights satisfying $\sum_{i=1}^N w_i =
1$, the quantity $\lambda$ satisfies $0 < \lambda < 1 $, while $\left\{ K_0( {\mathbf
x};{\mathbf y}), K_1( {\mathbf
x};{\mathbf y}), ... \right\}  $ are a set of local proposal
distributions, or ``kernels''.  In words, the \PD\ in
equation~(\ref{eq:bankpd}) says:
\begin{quote}
With probability $(1-\lambda)$ propose a local Metropolis step of the usual
kind, i.e.\ ``close'' to the last point in the chain.  With
probability $\lambda$, propose instead to move into the vicinity of
one of the number of ``banked'' points, chosen with weight $w_i$ from
within the bank.
\end{quote}
In practice, it may be convenient to take all of the kernel
distributions to be identical and equal to $K( {\mathbf x};{\mathbf
y})$, and all weights $w_i$ to be equal and identical to $1/N$.  We
shall use such Kernels and weights in the numerical examples studied
in this paper. Having made these choices, it will be seen that there
is only one free parameter, $\lambda$, in equation~(\ref{eq:bankpd})
-- the probability of making a ``bank'' step.  As $\lambda\rightarrow
0$ the \PD\ of equation~(\ref{eq:bankpd}) reverts to ordinary local
Metropolis behaviour, and no use is made of the ``banked''
information.  As $\lambda\rightarrow 1$ local behaviour is lost and
the sampler behaves more and more like an importance sampler.  It has
long been known that, at least in large numbers of dimensions, local
Metropolis usually fares better than importance sampling.  In practice
we will therefore want to use values of $\lambda$ that are close to
$0$, so as to favour exploration via the usual local random walk, but
not so close to $0$ as to make non-local steps using the bank
impossible.  Though the most efficient values of $\lambda$ will vary
on a case-by-case basis, we find empirically that $\lambda \sim 0.1$
is a good first value to try, not only for the toy problems considered
later, but also in real-world problems \cite{lester-allanach-prior}.

\subsection{Summary of the bank sampling algorithm}
\begin{itemize}
\item
When a new point ${\mathbf x}^{(t+1)}$ is to be sampled, a proposal
${\bf x'}$ is first made as follows.
With probability $(1-\lambda)$ the proposal ${\bf x'}$ is chosen in the vicinity of
the last point ${\mathbf x}^{(t)}$ using a local proposal kernel $K_0(
{\mathbf x'};{\mathbf x}^{(t)})$, such as a scaled normal distribution.
With complementary probability $\lambda$ the proposal must instead be chosen from the
vicinity of one of the banked points ${\mathbf y}^{(i)}$ with weight
$w_i$, using a possibly point-dependent proposal kernel $K_i( {\mathbf
x'};{\mathbf y}^{(i)} )$.
\item
The value
\begin{equation}
\rho =
\frac{{\mathcal L}({\bf x'})}{{\mathcal L}{({\bf x}^{(t)} )}} \cdot
\frac
{
(1-\lambda) K_0( {\mathbf x}^{(t)};{\mathbf x'}       ) + 
\lambda \sum_{i=1}^N {
w_i K_i(     {\mathbf x}^{(t)};{\mathbf y}^{(i)}     )}
}
{
(1-\lambda) K_0( {\mathbf x'};{\mathbf x}^{(t)}       ) + 
\lambda \sum_{i=1}^N {
w_i K_i(     {\mathbf x'};{\mathbf y}^{(i)}     )}
}
\label{eq:formorons}
\end{equation}
is then computed.
\item
The proposal ${\bf x'}$ is accepted as the new sample ${\mathbf
  x}^{(t+1)}$ with probability $\min[\rho,1]$.  If ${\bf x'}$ is
  rejected, then the previous point ${\mathbf x}^{(t)}$ becomes
  ${\mathbf x}^{(t+1)}$ instead.
\end{itemize}

\subsection{Worst case costs of bank sampling}

What is the cost of applying the bank sampler to a situation in which
it is ill suited?  How costly is it to use the bank sampling
algorithm when the set of ``clues'' are no good?  The worst case cost
in these situations it typically only 10\% more than standard \MHA.\footnote{We
discuss a realistic example of the cost of overhead of obtaining the bank
samples in section~\ref{sec:mde}.}

This is because for a fraction $(1-\lambda)$ of the time the bank
sampler is doing standard \MHA\ anyway.  Even if no bank-based
proposal is ever accepted, the remaining part of the algorithm will
carry on regardless for the proportion $(1-\lambda)$ of the time in
which it is used.  Since $\lambda$ will typically be chosen to be
$\sim 0.1$, this means that a maximally poor choice of bank would be
expected to lead to an order 10\% increase in computing
time.\footnote{We note that the preceding statements assume, in line
with most realistic scenarios in particle physics, that the cost of
evaluating the ``banked'' $Q$-factors in equation~(\ref{eq:formorons})
is negligible in comparison to the cost of evaluating the target
density ${{\mathcal L}({\bf x})}$.  In situations in which target
densities are cheaper than proposal functions, the cost could be
higher than 10\%.}

\section{Examples}

\subsection{Two-dimensional toy problem}

\label{sec:two-d-toy}

\begin{figure}
\epsfig{file=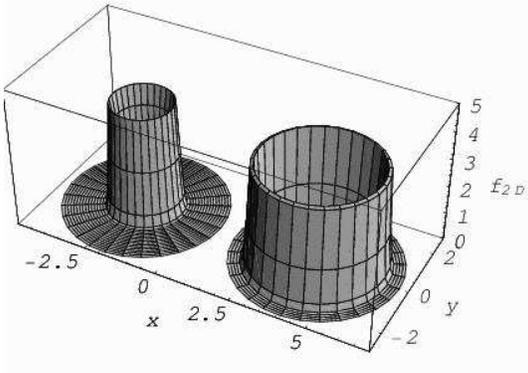, width=0.698 \wth}
\caption{The example two-dimensional
  distribution $f_{2D}({\bf x})$ defined in
  equation~\ref{eq:twodfunc}. \label{fig:diagof2dfunc}}
\end{figure}

We demonstrate the performance of the Bank Sampler graphically with a
two-dimensional toy problem in which we choose the
target distribution to be:
\begin{equation}
f_{2D}({\bf x}) = 
\circf({\bf x}; c_1, r_1, w_1)+
\circf({\bf x}; c_2, r_2, w_2) \label{eq:twodfunc}
\end{equation}
where $c_1=(-2,0)$, $r_1=1$, $w_1=0.1$, $c_2(+4,0)$, $r_2=2$, $w_2=0.1$ and
\begin{equation}
\circf({\bf x};{\bf c},r,w) = \frac{1}{\sqrt{2 \pi w^2}}
\exp\left[-\frac{(|{\bf x}-{\bf c}|-r)^2}{2 w^2}\right].
\end{equation}
This toy distribution, depicted in figure~\ref{fig:diagof2dfunc},
represents two well separated rings, each of which has a narrow
gaussian profile.  This distribution is chosen as it has a mixture of
short scales and long scales typical of many beyond-the-Standard-Model
parameter space scans of the type that the Bank Sampler was designed
to cope with.

\begin{figure}
\ninegraphs
{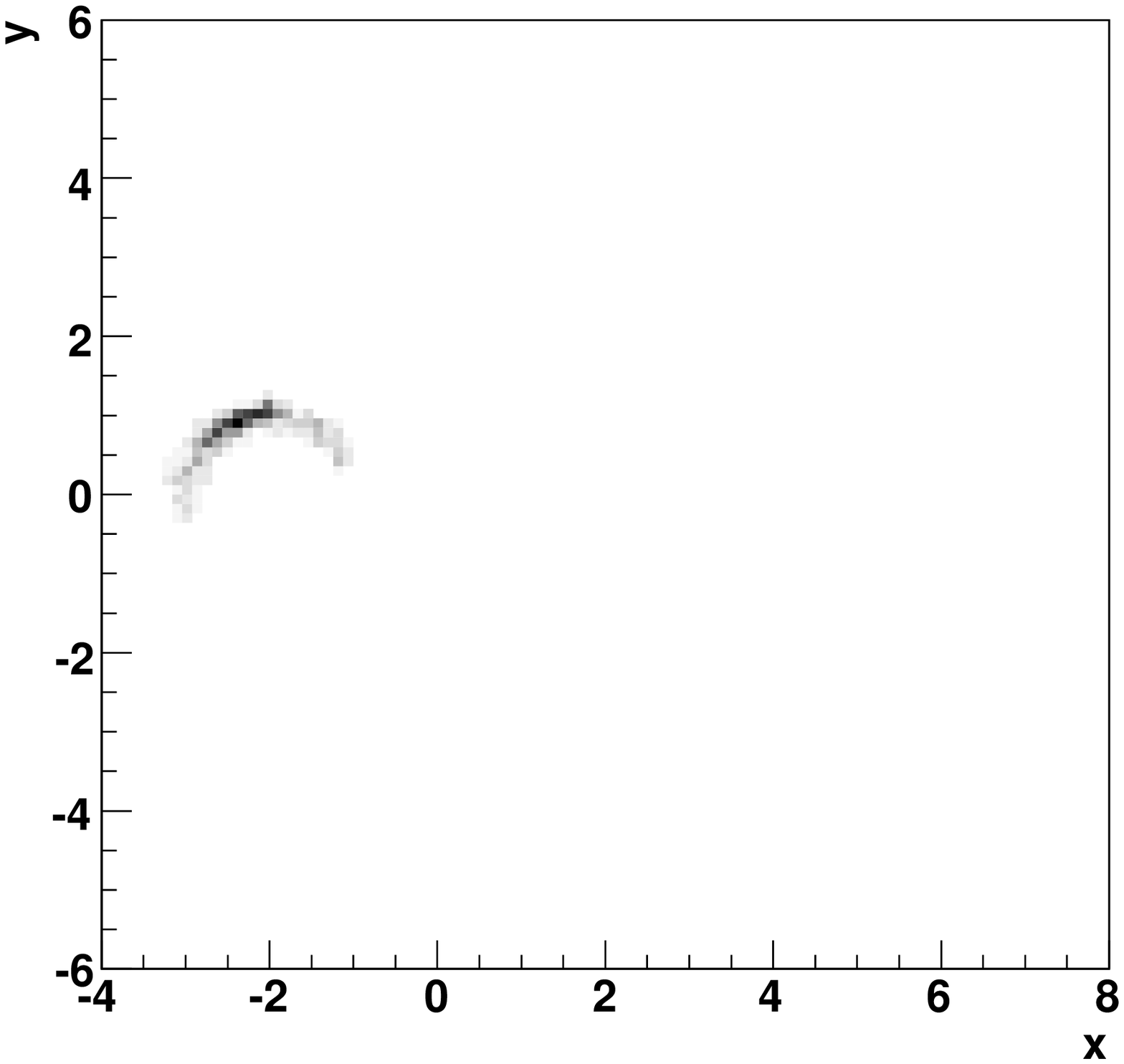}
{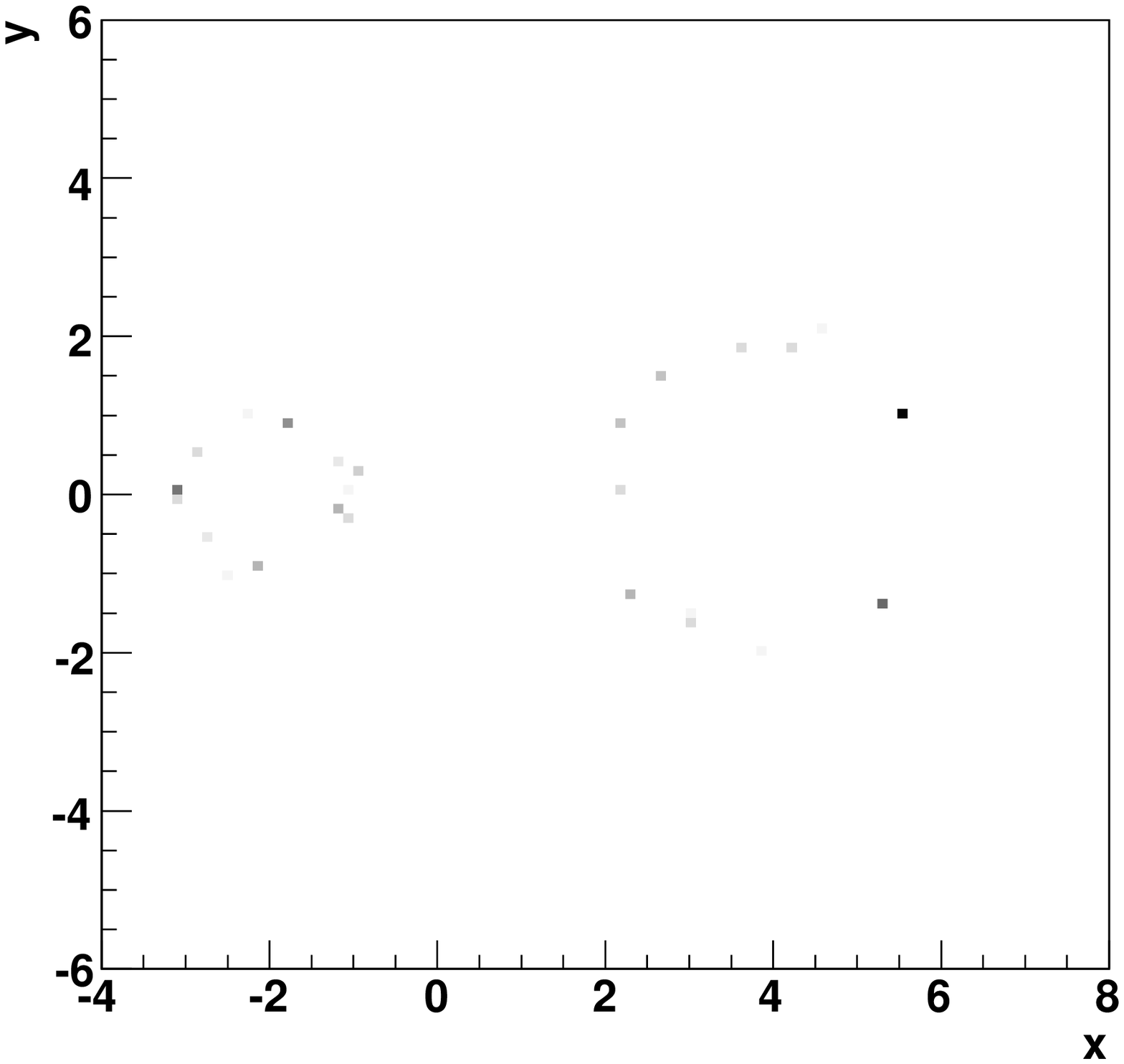}
{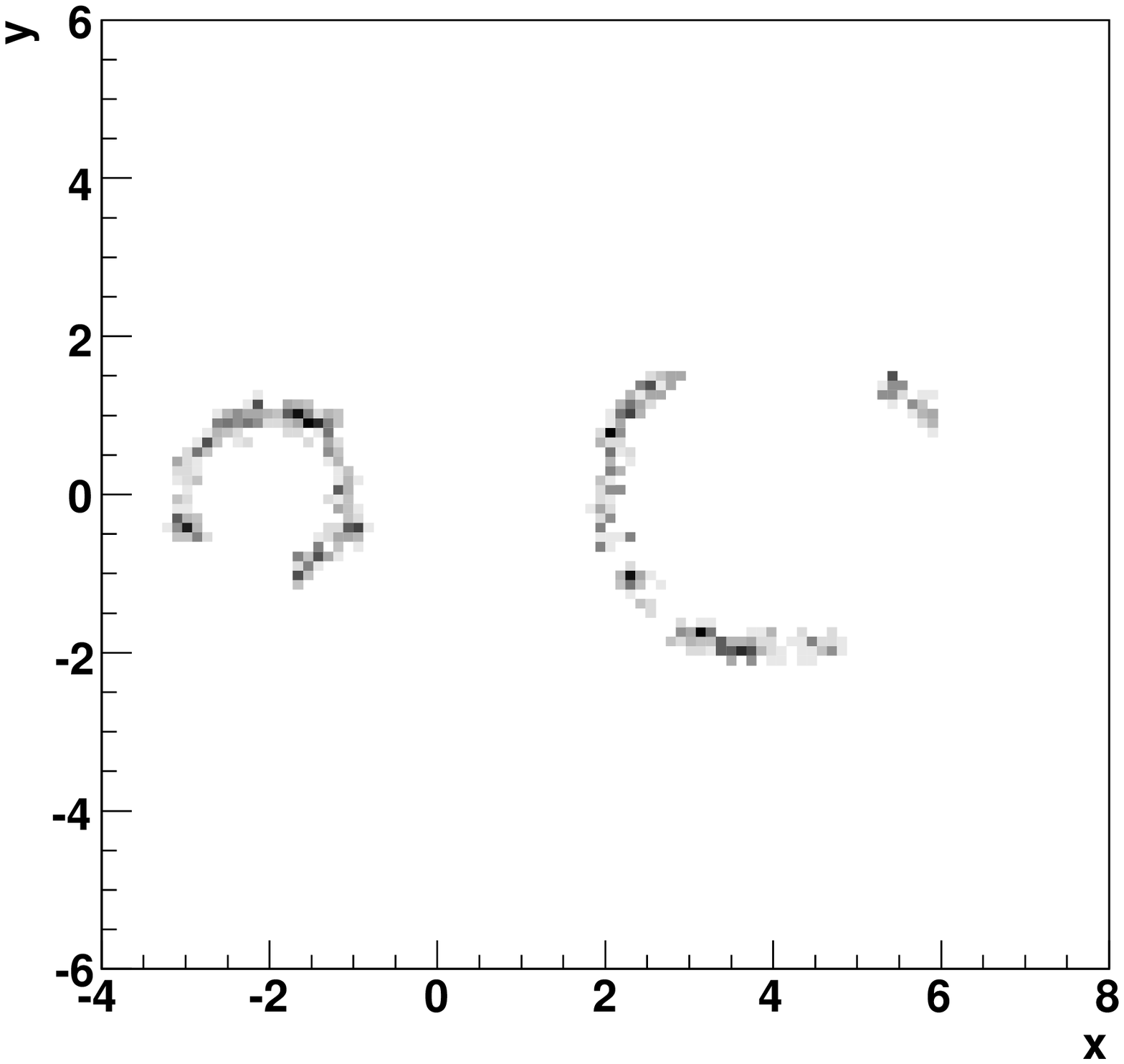}
{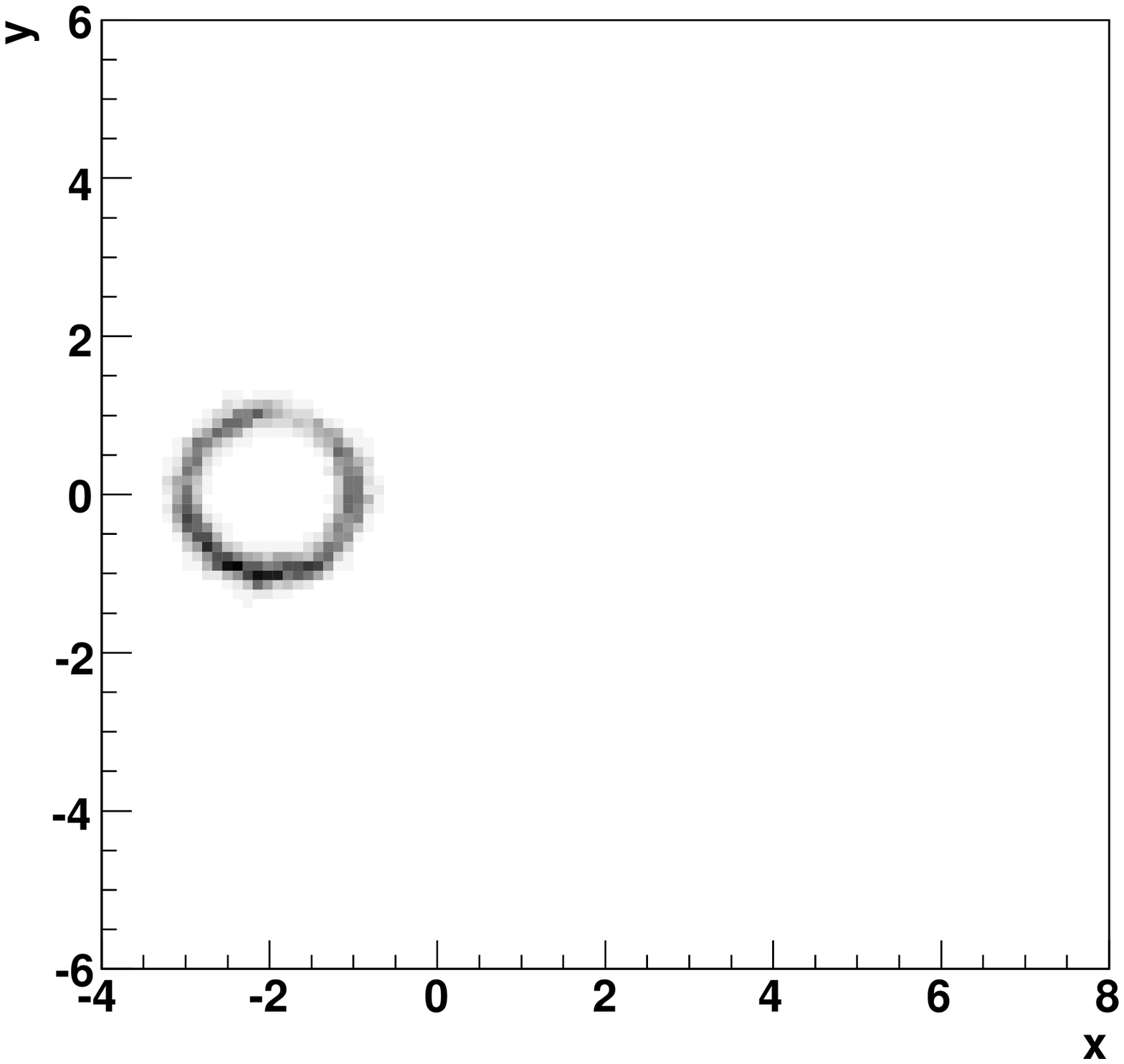}
{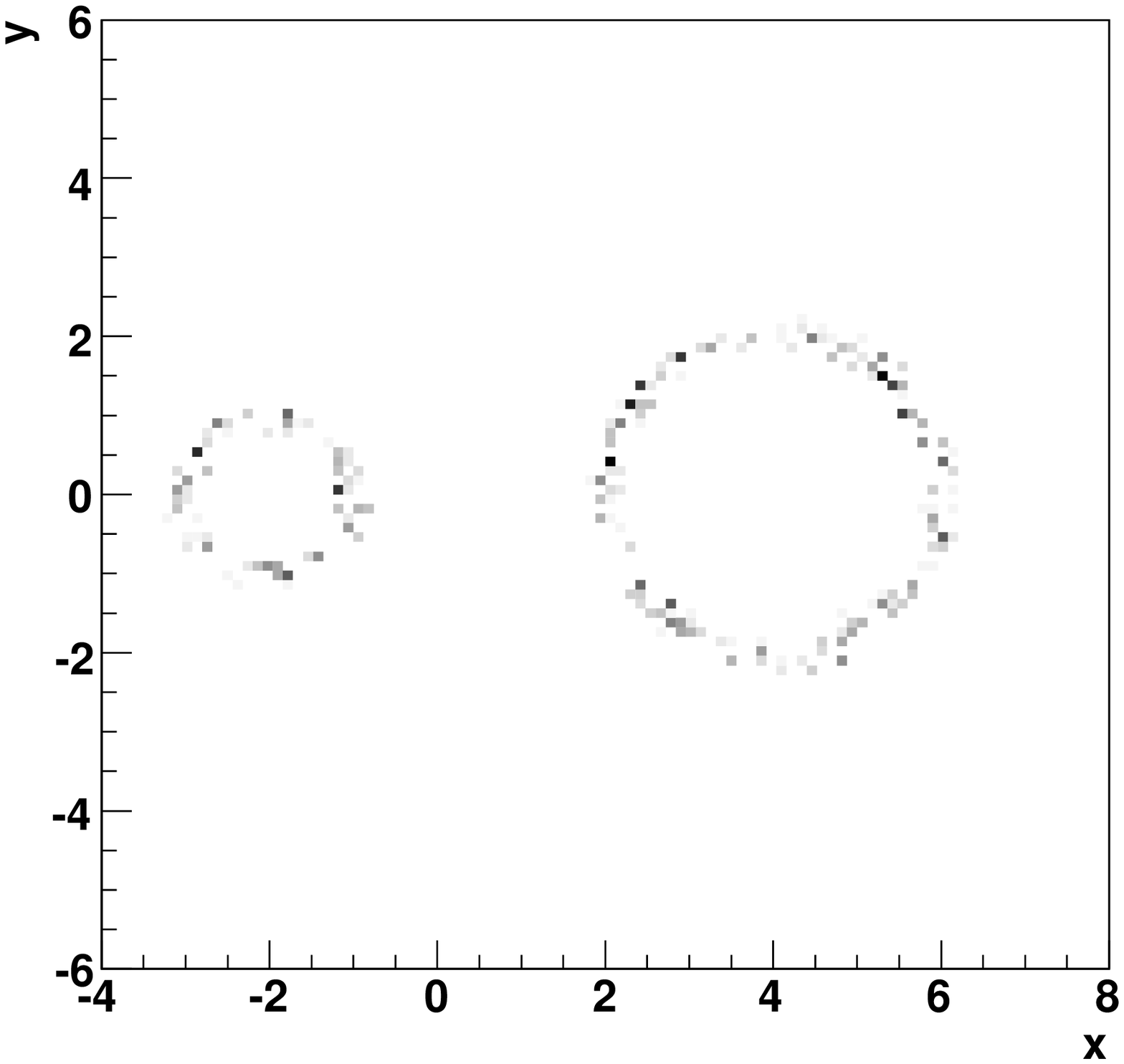}
{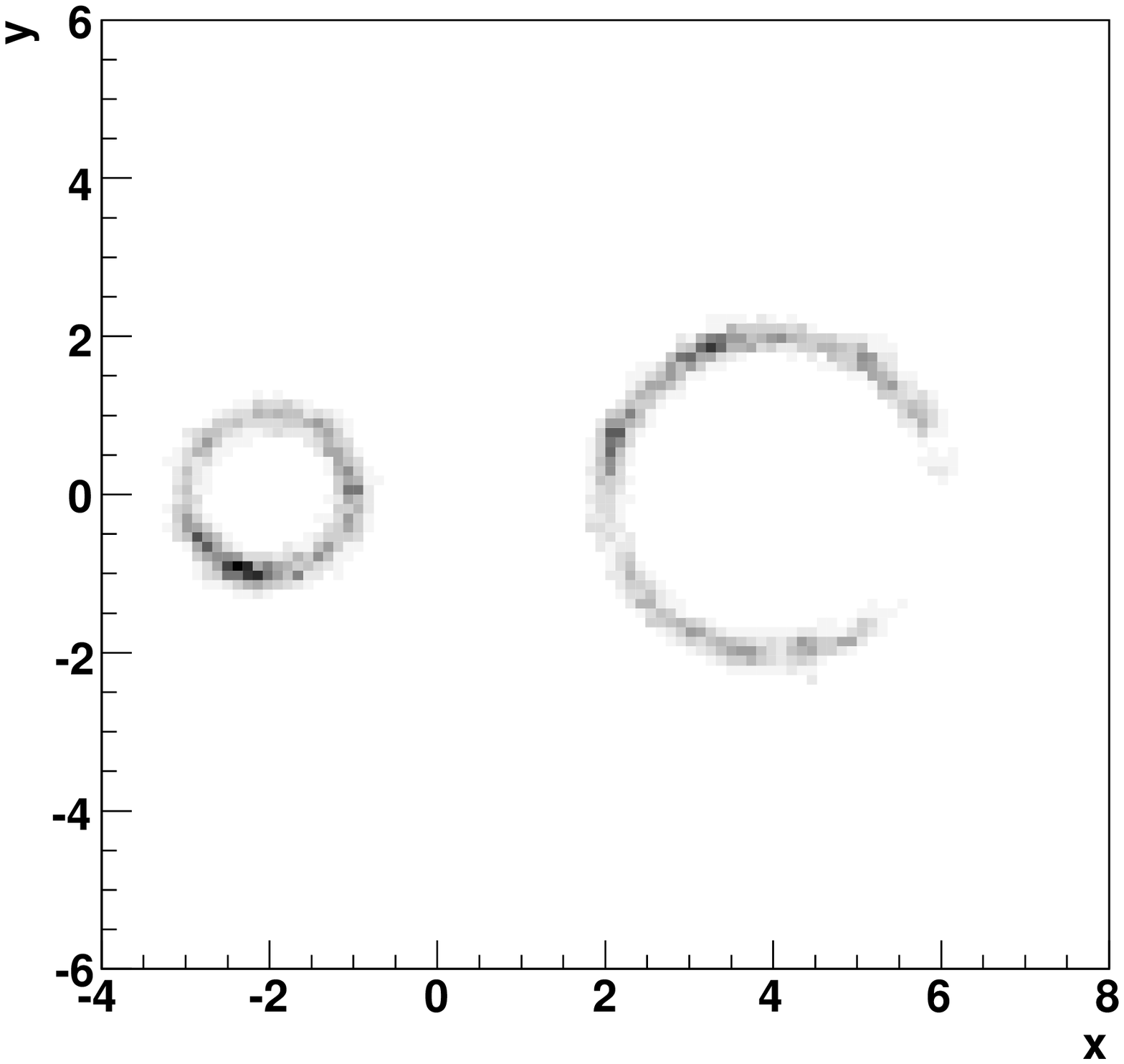}
{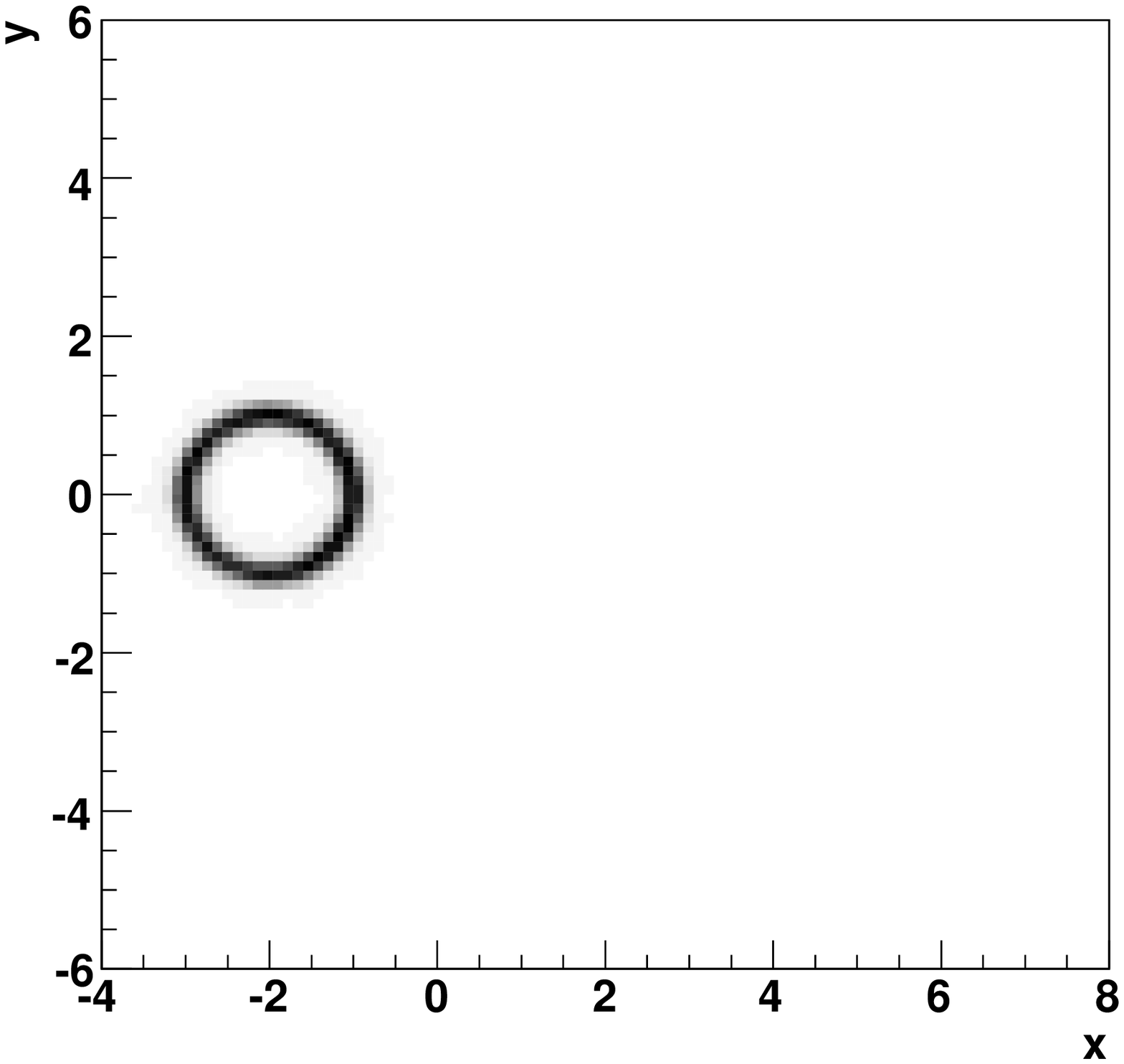}
{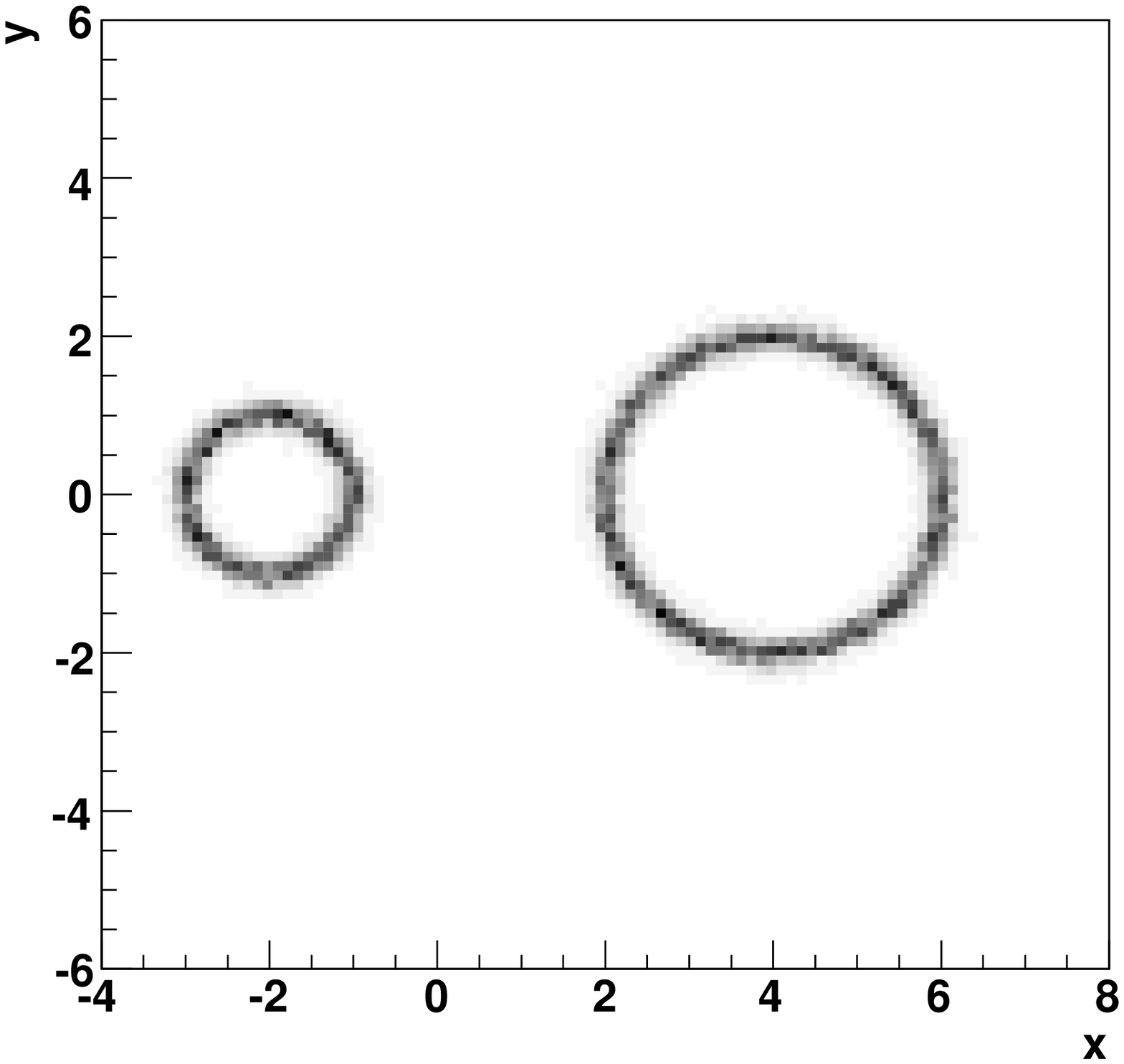}
{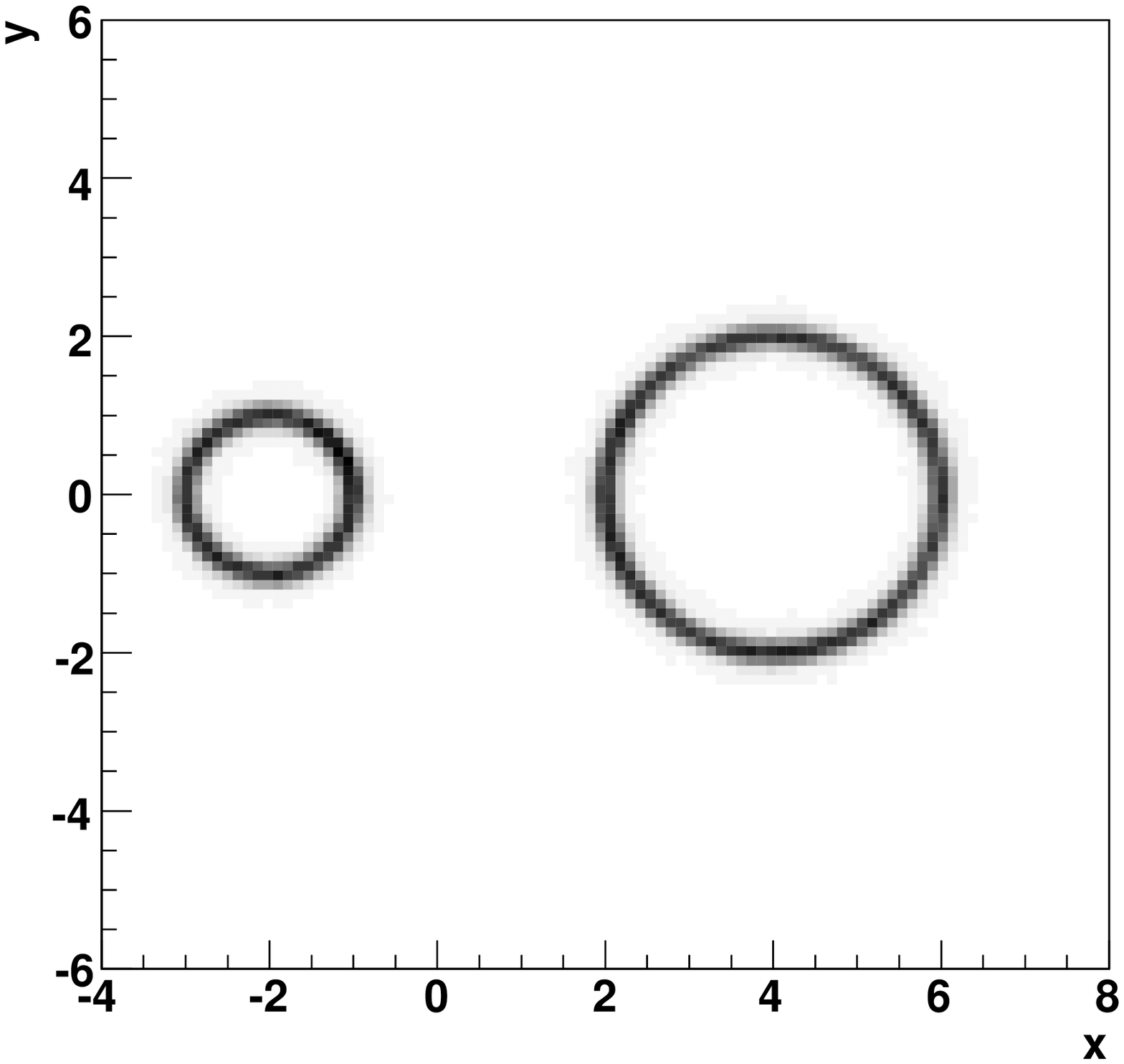}
\caption{2D example\label{fig:2D-circs}.  This figure compares the
performance of two metropolis samplers (``normal'' in the left hand
column, ``broad'' in the middle column) with
the performance of the ``bank sampler'' sampler (right hand column). 
Details of the samplers are given in the text.  The number of samples
increases in each successive row of the table: one-thousand samples in
the top row, ten-thousand samples in the second row, and one-million
samples in the bottom row.}
\end{figure}

Figure~\ref{fig:2D-circs} compares the performance of two ordinary
Metropolis samplers (left hand and middle columns) with the
performance of a ``bank sampler'' sampler (right hand column) using
twenty banked points.  The bank sampler is defined by
equation~(\ref{eq:bankpd}) with $\lambda=0.1$, and equal weights $w_i$
and a common kernel.  Ten bank points were given on the large circle
and ten on the small circle, each at nominal distance from circle
centre, but with angles chosen uniformly from $[0, 2 \pi]$.  The
common kernel was a 2-dimensional Gaussian distribution of width
$w=0.1$, i.e.~$K({\bf x};{\bf y})=1/(2 \pi w^2) \exp[-({\bf x}-{\bf
y})^2/(2 w^2)]$.  The two implementations of the ordinary Metropolis
algorithm used to provide the comparison samples were derived from the
bank sampler above by the setting of $\lambda = 0$ in
equation~(\ref{eq:bankpd}).  The first (termed the ``normal''
Metropolis algorithm) used a kernel width $w=0.1$ identical to that of
the bank sampler.  The second (termed the ``broad'' Metropolis
algorithm) used a kernel width $w=6$ equal to the centre-to-centre
separation of the circles.  No burn in time is associated with these
samplings as the initial point was always chosen in the typical set.

The ``normal'' Metropolis sampler is unable to move over to the
right-hand circle centred on $(4,0)$ in one million steps as
Figure~\ref{fig:2D-circs}(g) shows.  The bank sampler is seen to keep
the two circles in equilibrium from as few as 1000 samples as seen
in Figures~\ref{fig:2D-circs}(f) and \ref{fig:2D-circs}(i).

In situations similar to those for which the bank sampler was created
(for more details see section~\ref{sec:mde}) one has most of the
probability mass concentrated in an extended multi-dimensional sheet
or hypersurface with very small intrisic thickness.  This hypersurface
is usually not simply connected, and is often highly curved or
folded.  In the authors' experience of these distributions, the best
simple (i.e.\ single kernel) Metropolis implementations have tended to
use proposal distributions whose length scales were all small -- of
the order of the thickness of the sheet.  In these cases, creating
thicker-tailed Cauchy or other non-gaussian distributions did not
usually assist progress around the hypersurface, and indeed always reduced
the sampling efficiency by increasing the proportion of rejected
proposals.\footnote{Due to the vanishing probability of stepping into
a portion of the thin hypersurface somewhere distant in the space.}
For the above reasons, we believe that the ``normal'' Metropolis
algorithm (which shares the length scale of the proposal used in the
bank sampler) is the more appropriate algorithm against which to
compare the bank sampler.  It represents what you might implement if
you knew you had a thin sheet to wander around, and didn't know where
any other sheets were.

\clearpage

\begin{figure}
\ninegraphs
{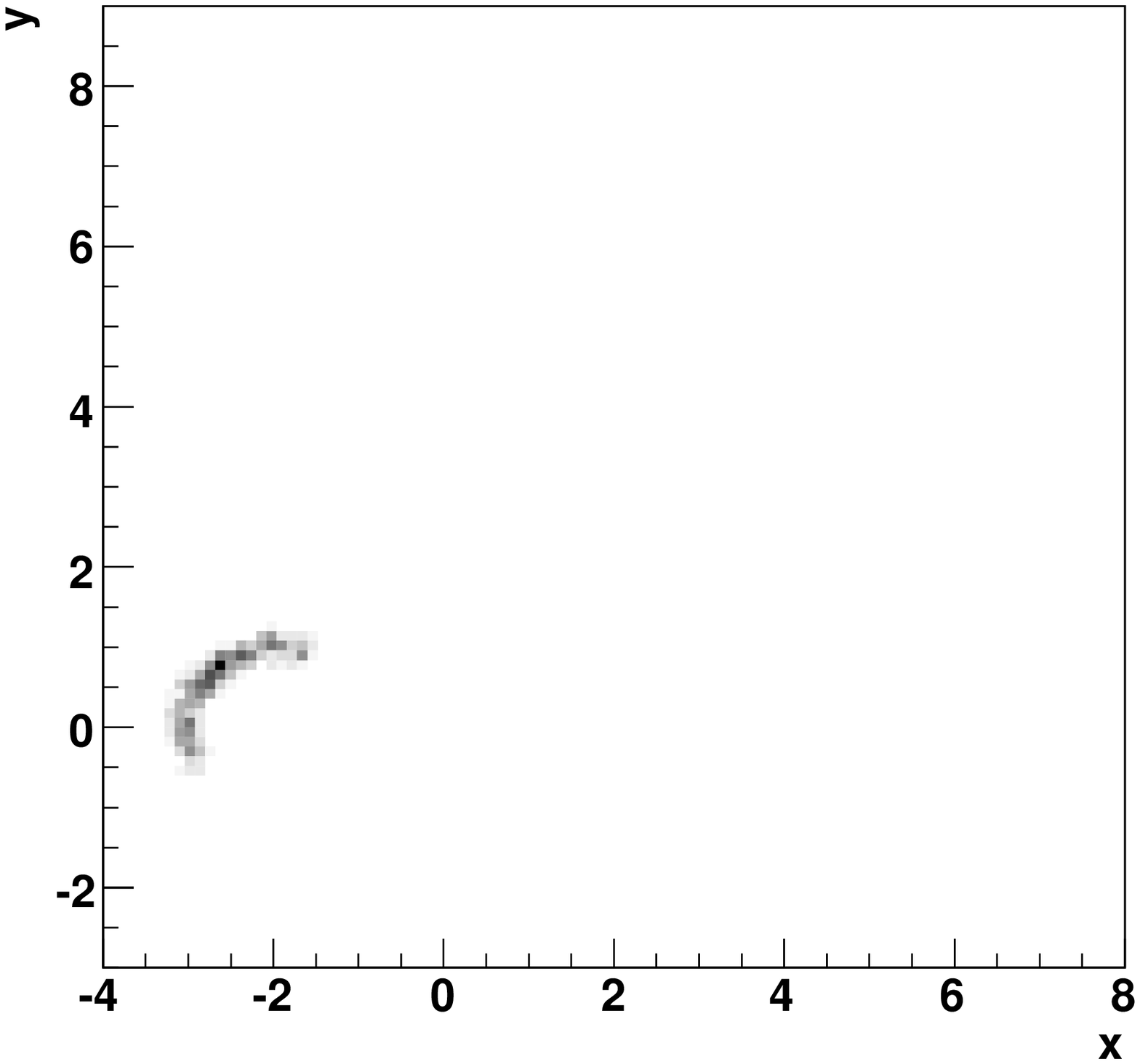}
{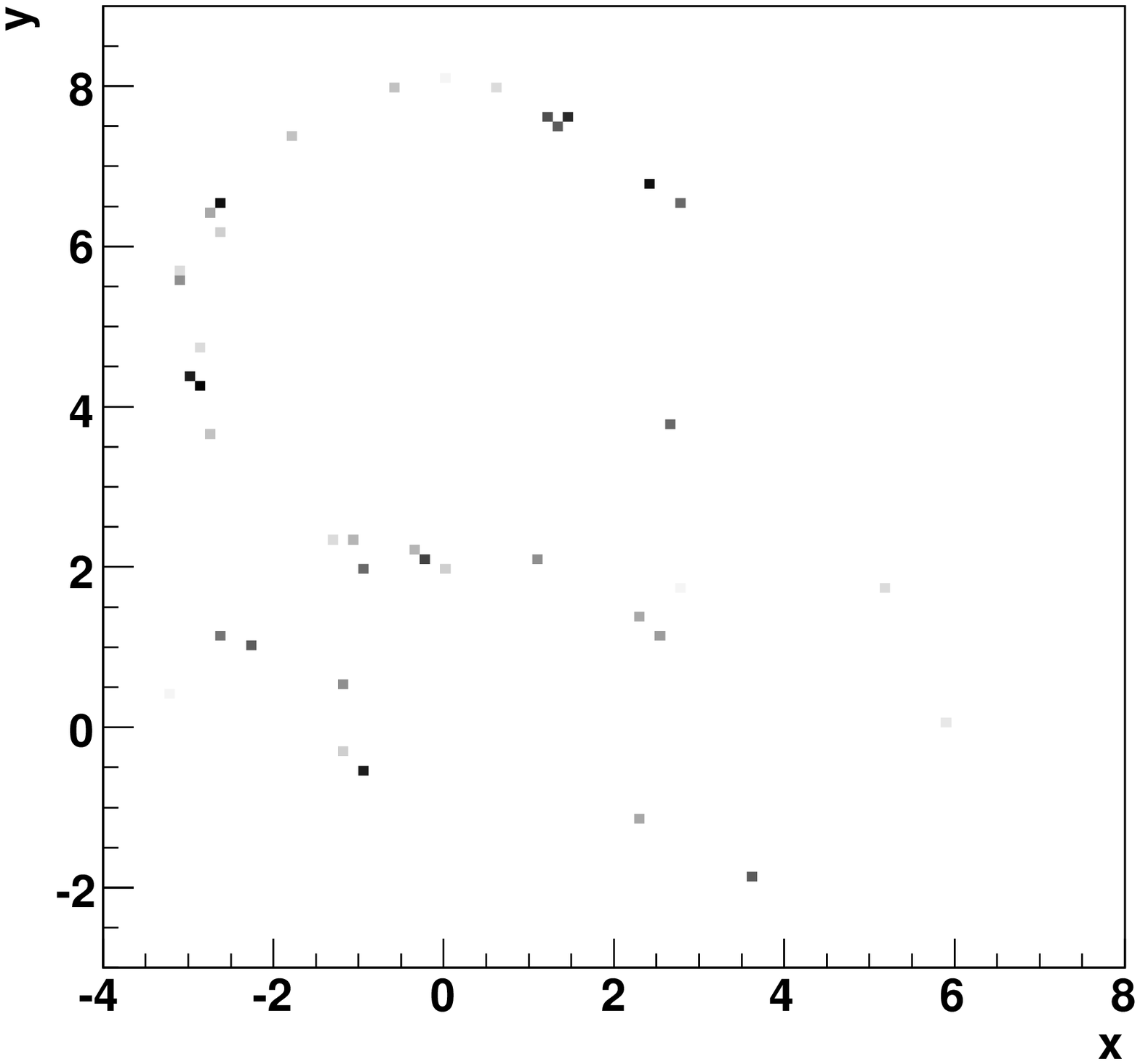}
{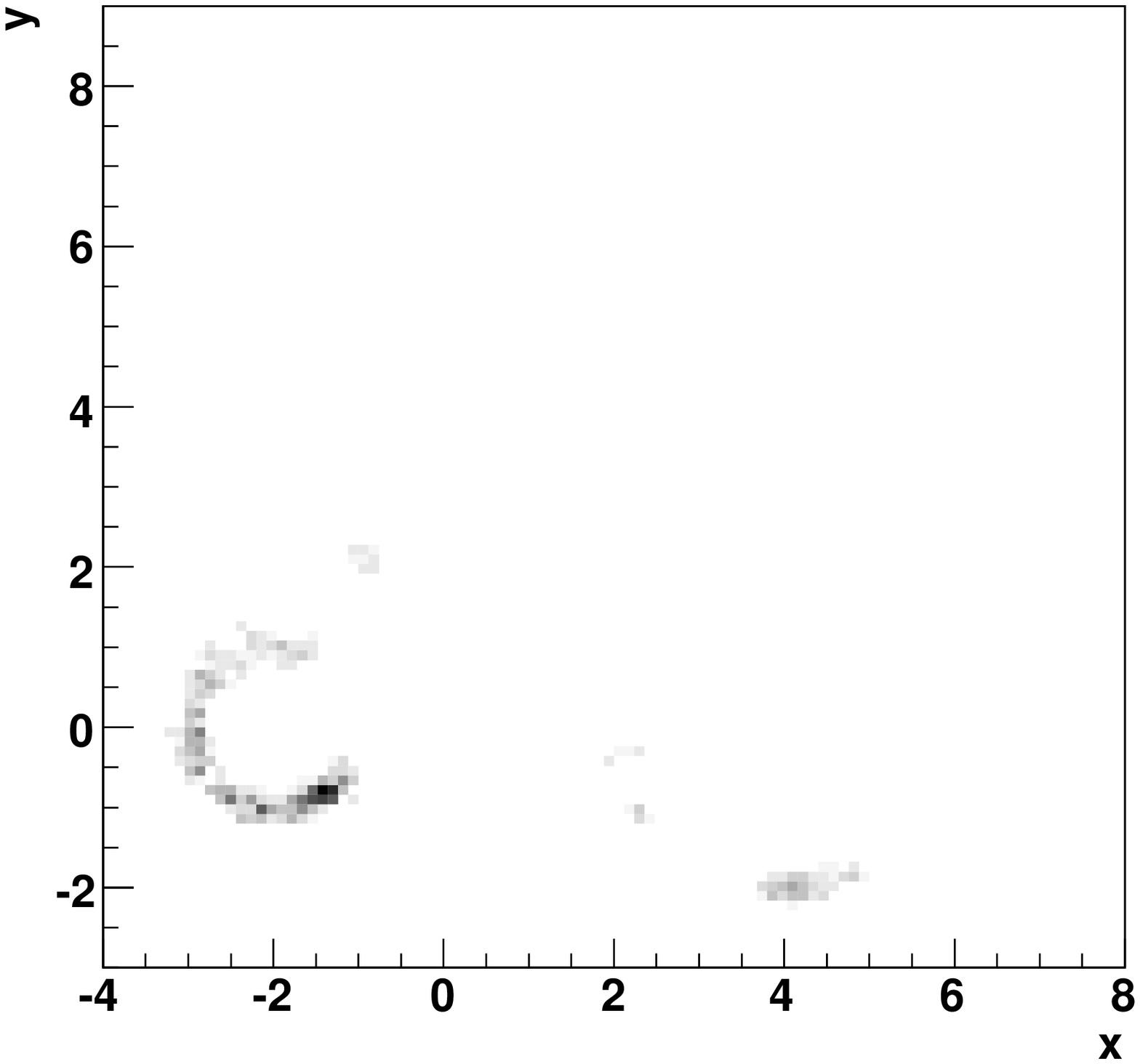}
{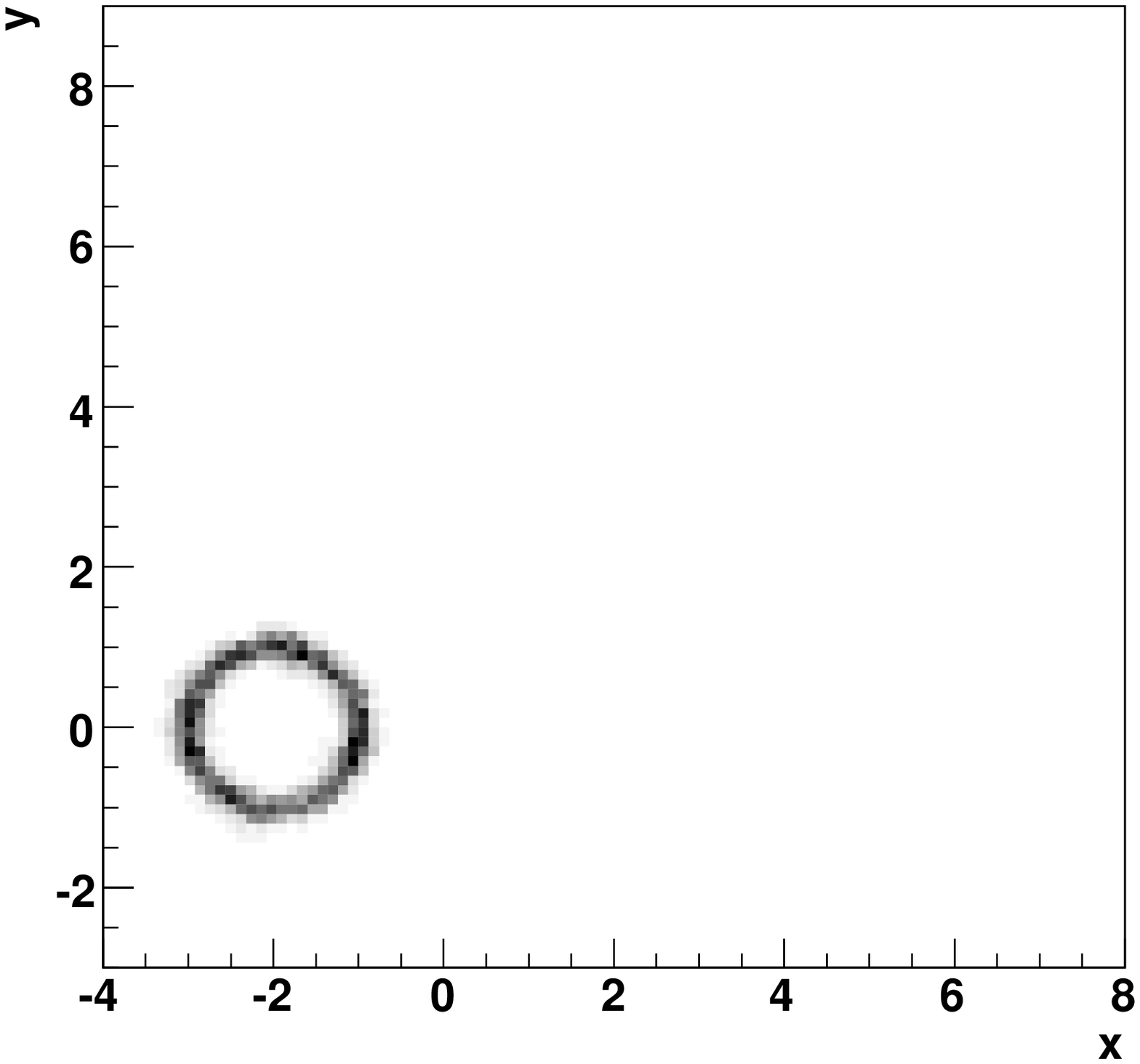}
{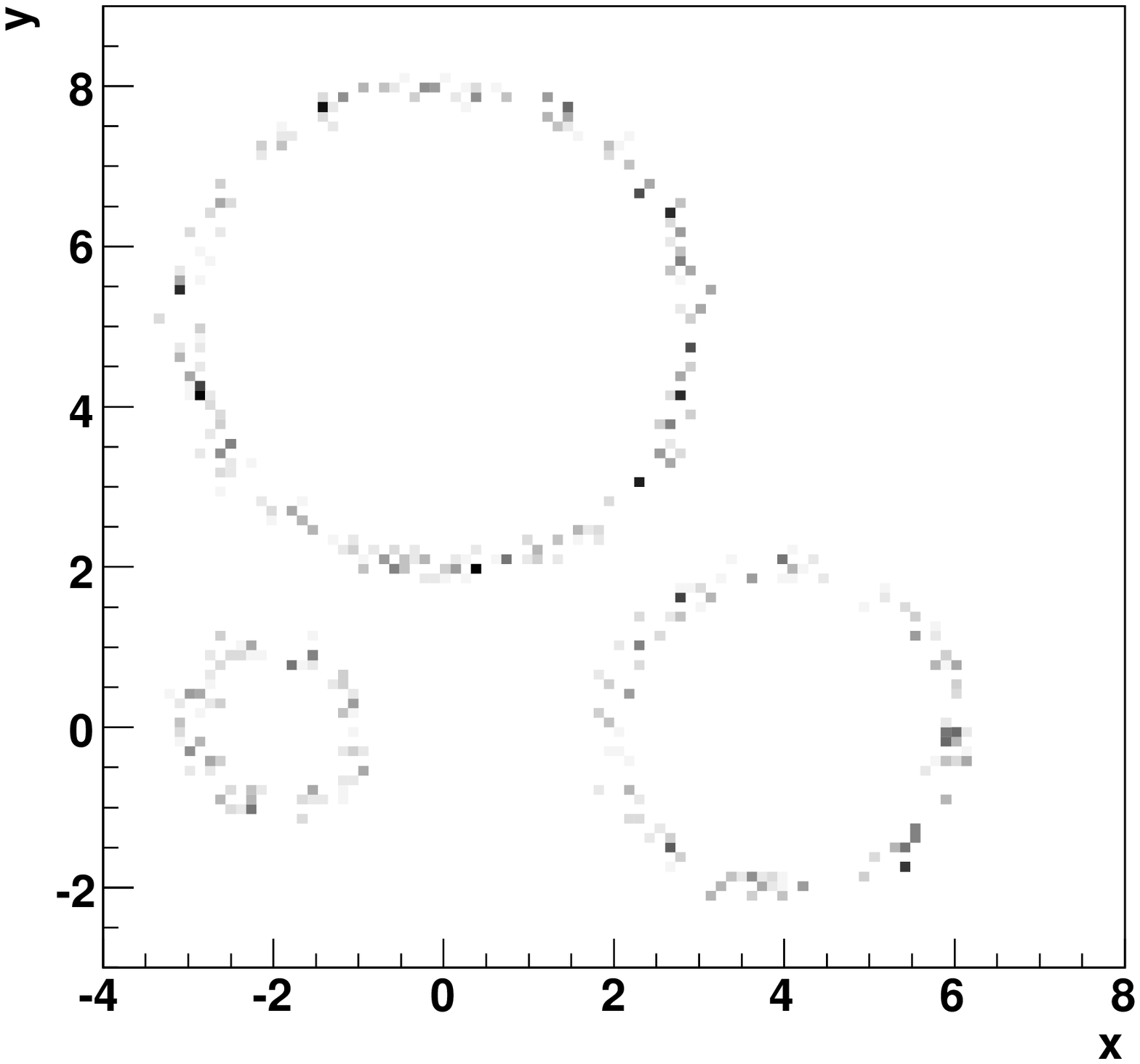}
{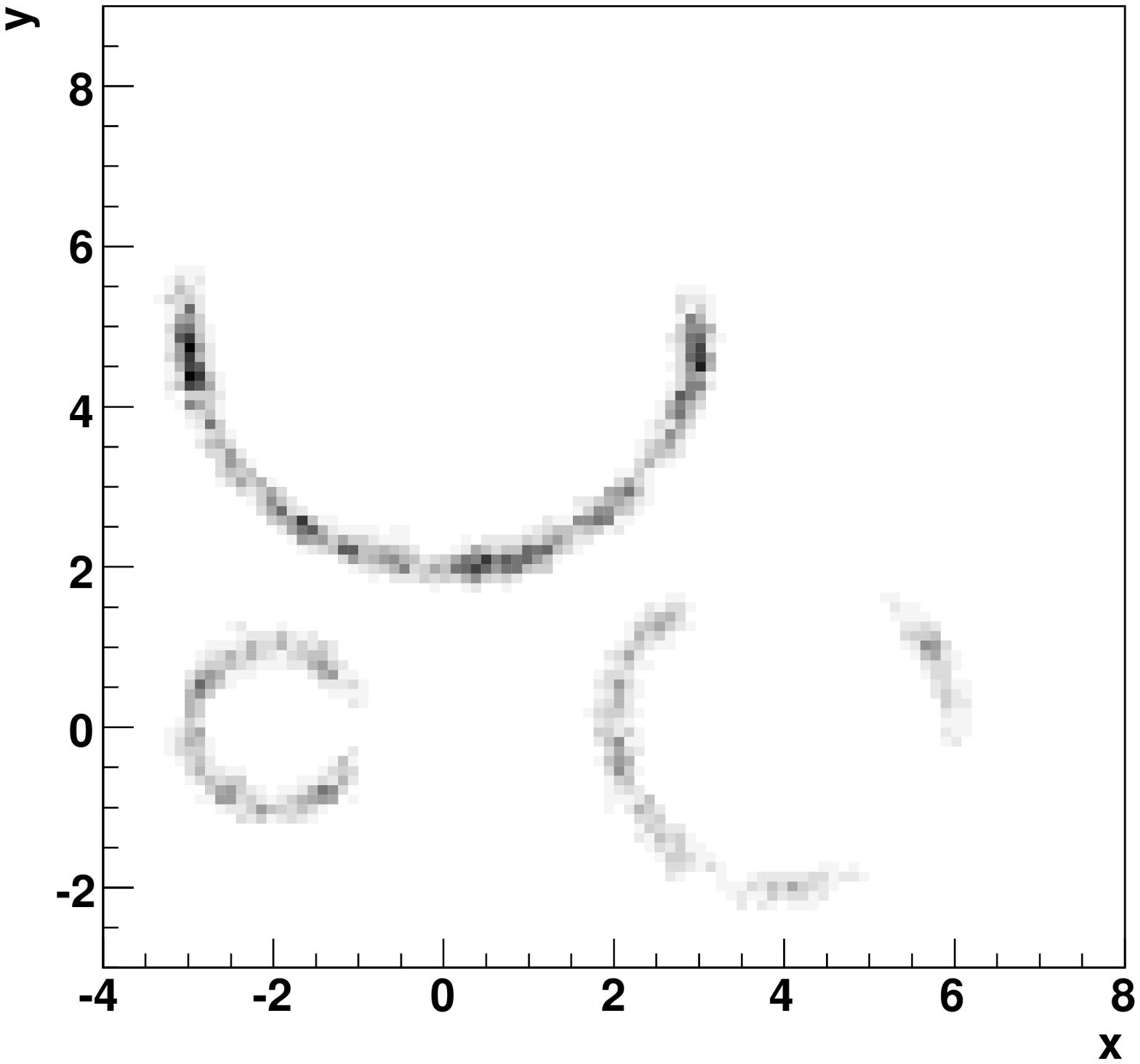}
{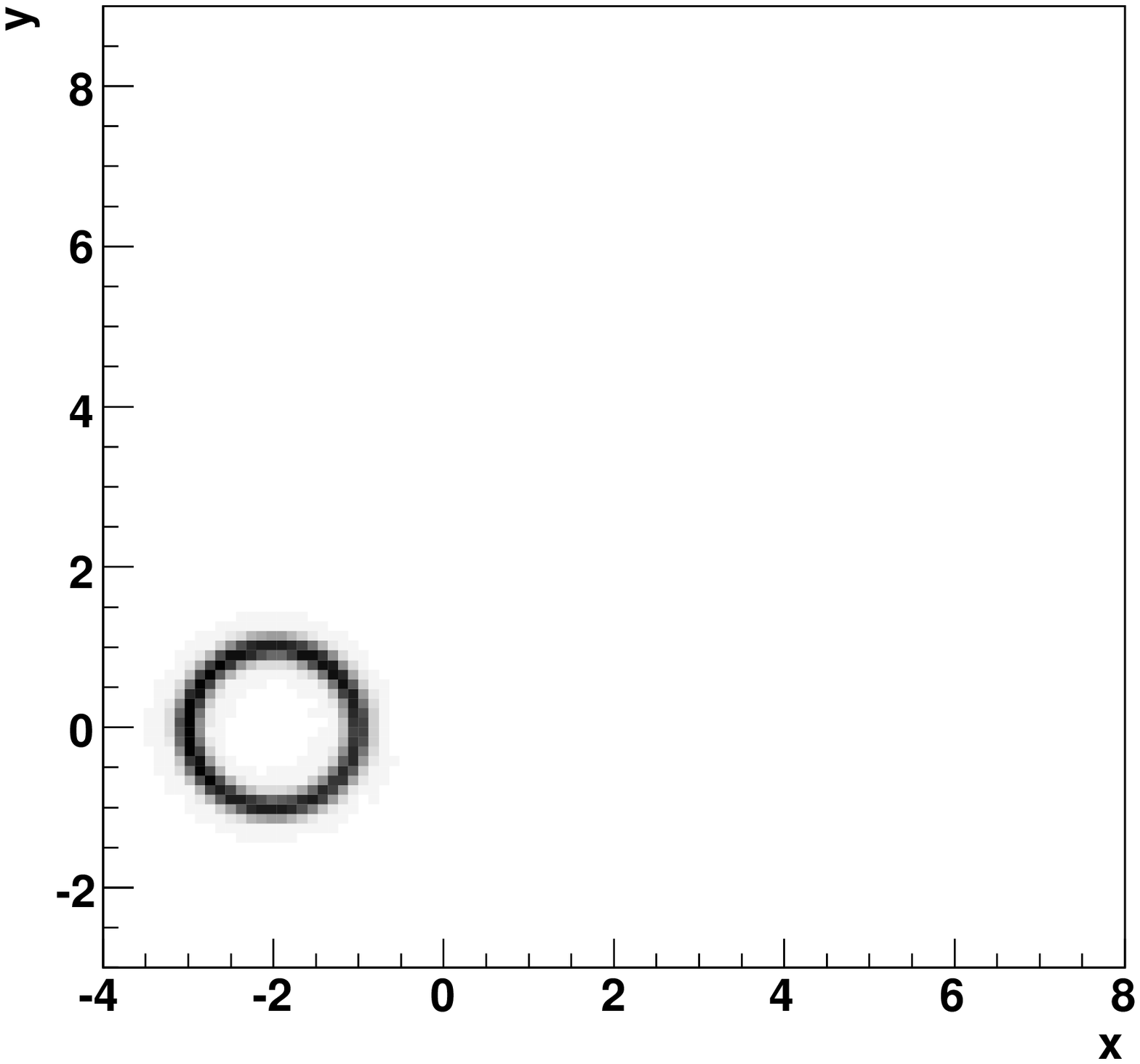}
{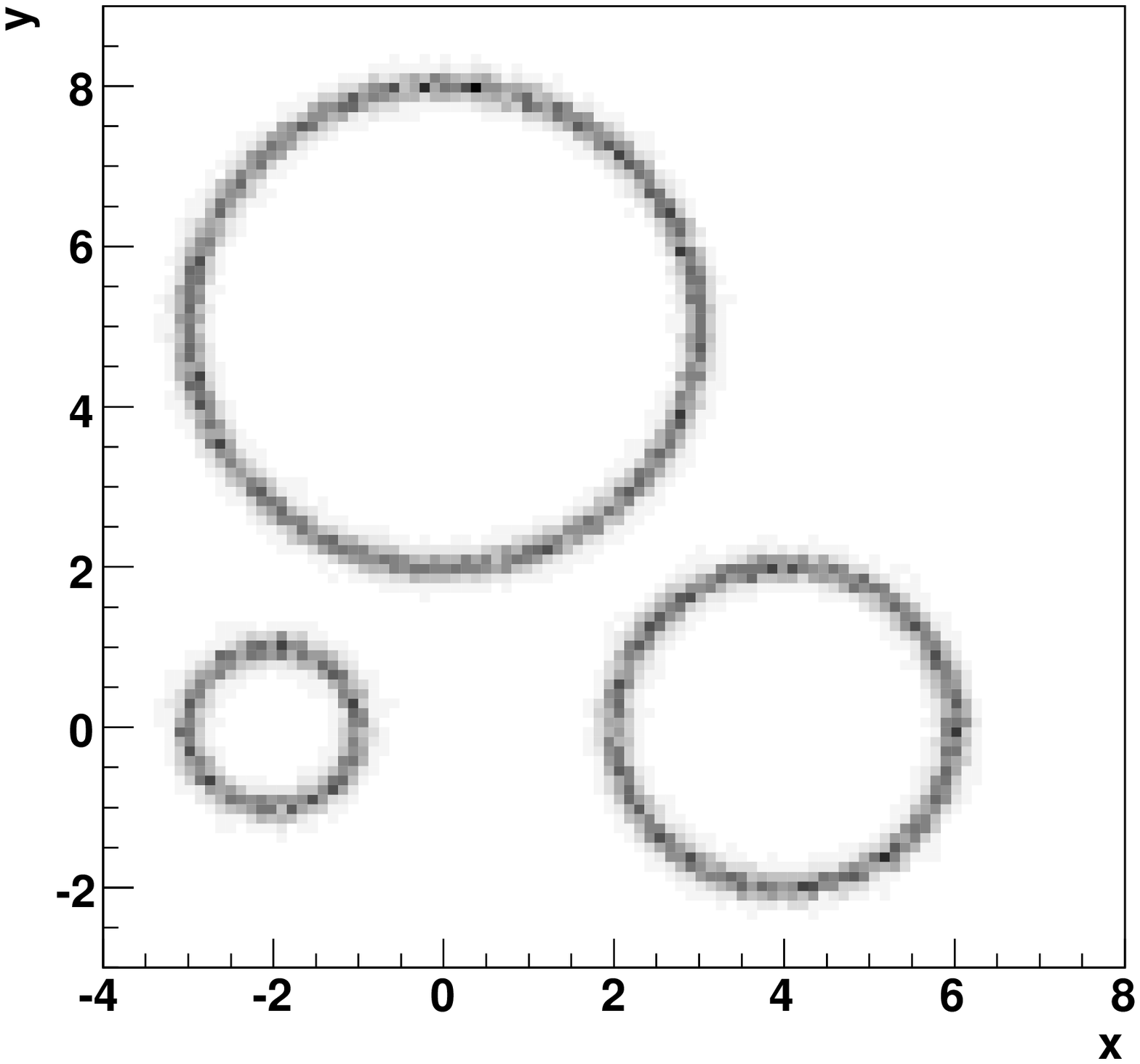}
{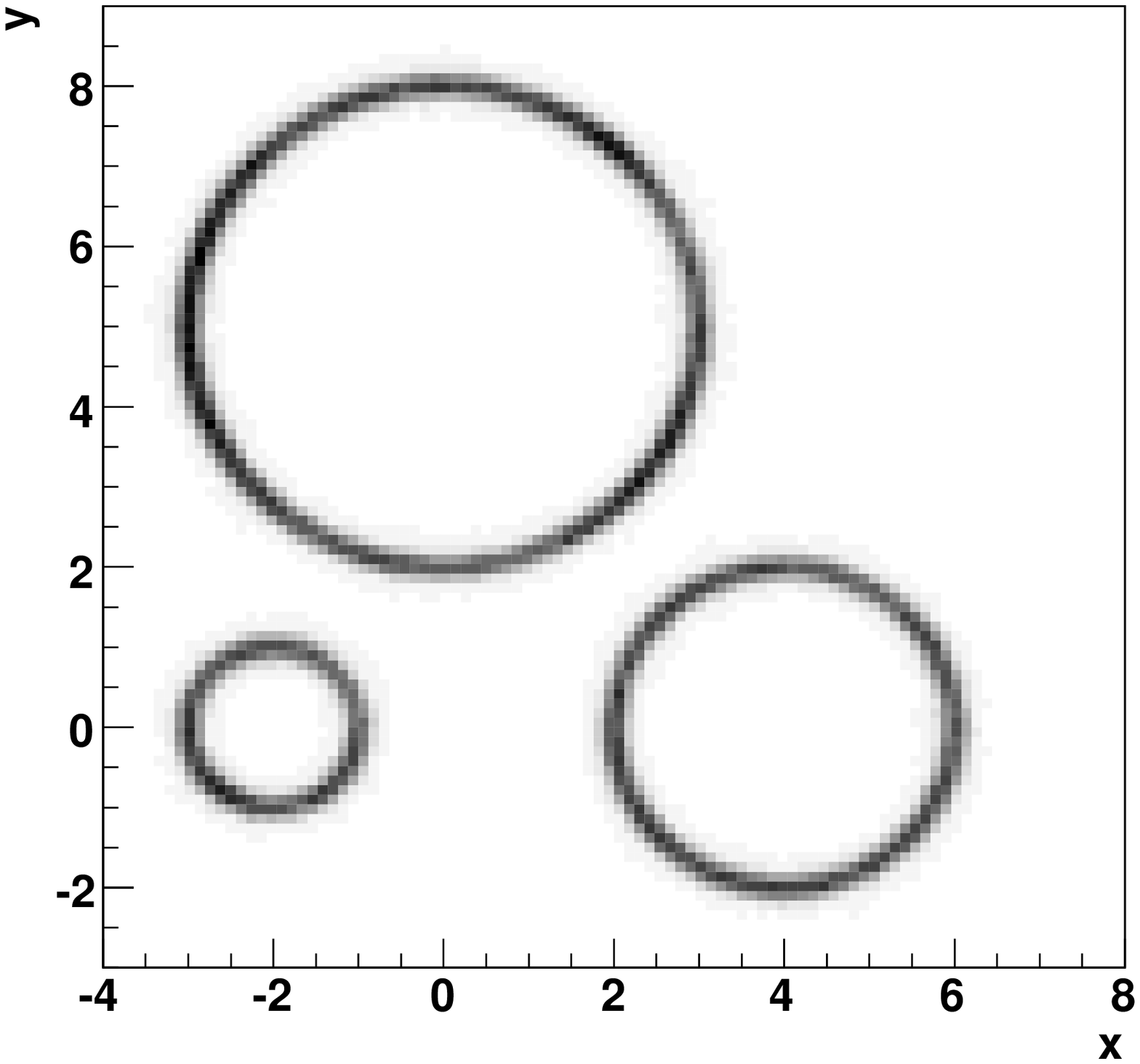}
\caption{Another 2D example, mostly identical to that shown in figure
  ~\ref{fig:2D-circs} except for the following two changes: (1) an
  additional ``large'' circle has been added, and (2) {\bf the
  distribution of the ``bank'' of clues has been deliberately skewed}:
  the small circle has been given 10 clues, the medium circle has been
  given 5 clues, and the largest circle has been given only 1
  clue. \label{fig:2D-circs-THREE-CIRCS} The meanings of the rows and
  columns are the same as in figure~\ref{fig:2D-circs}.}
\end{figure}

\begin{figure}
\ninegraphs
{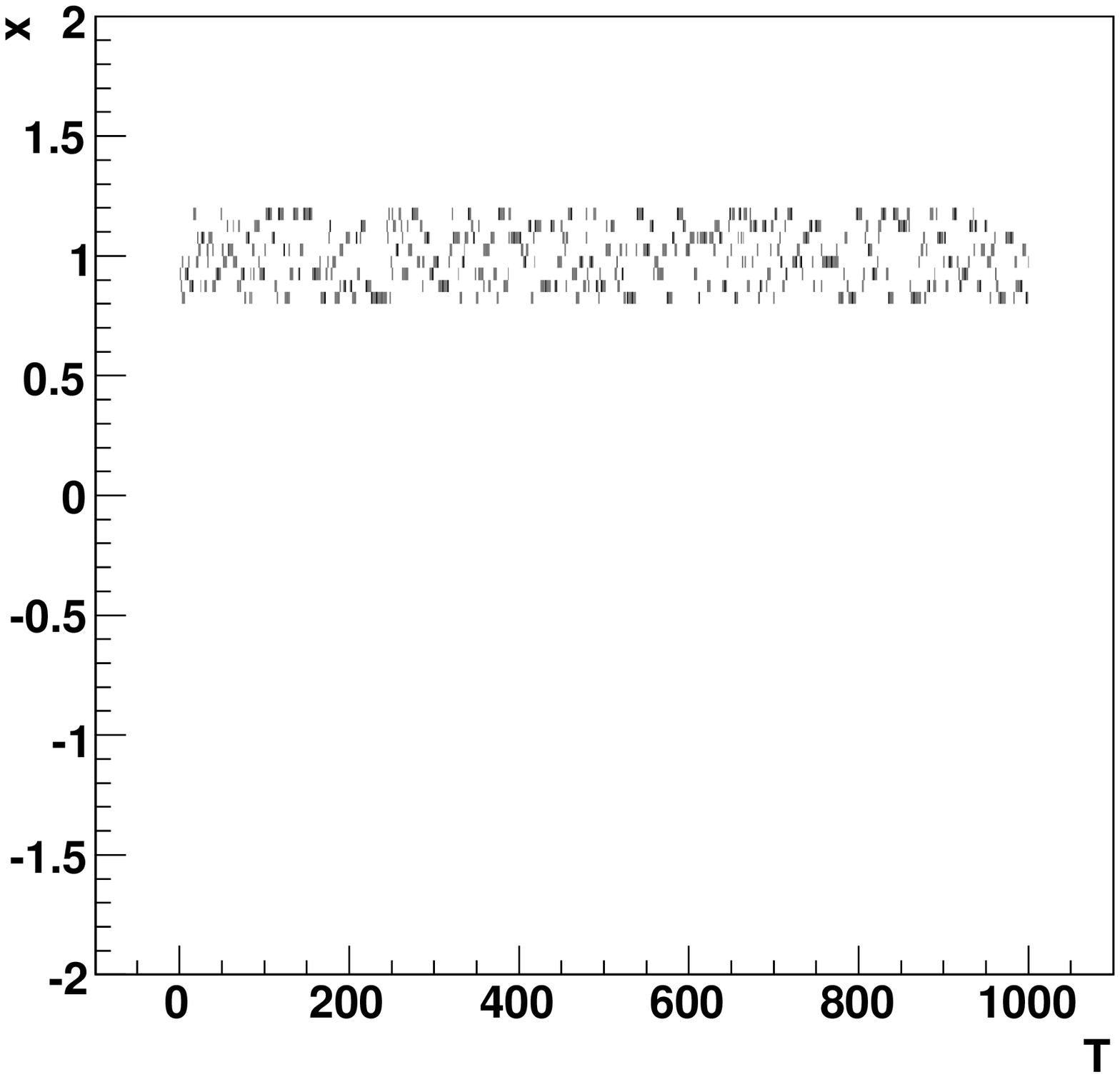}
{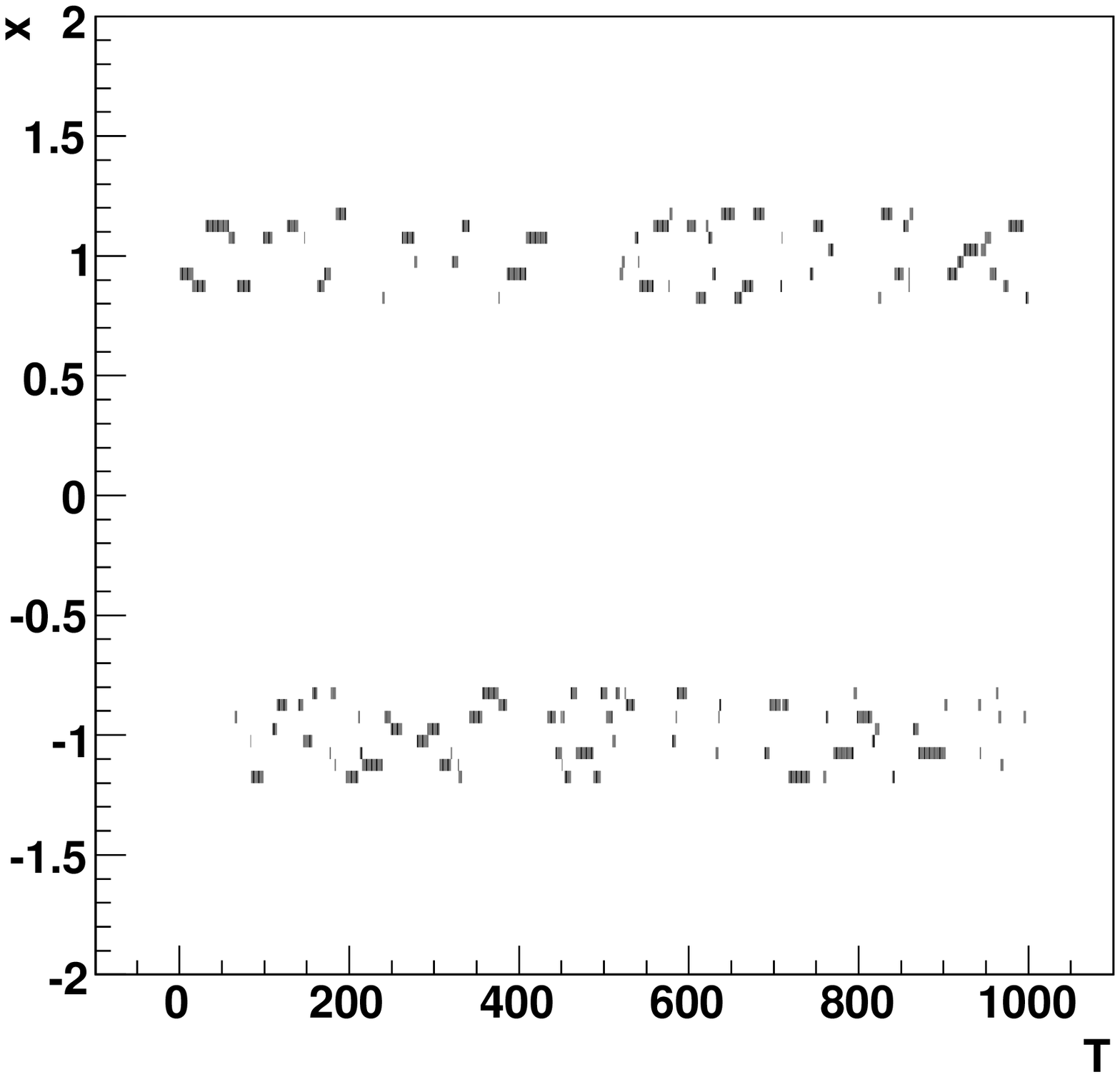}
{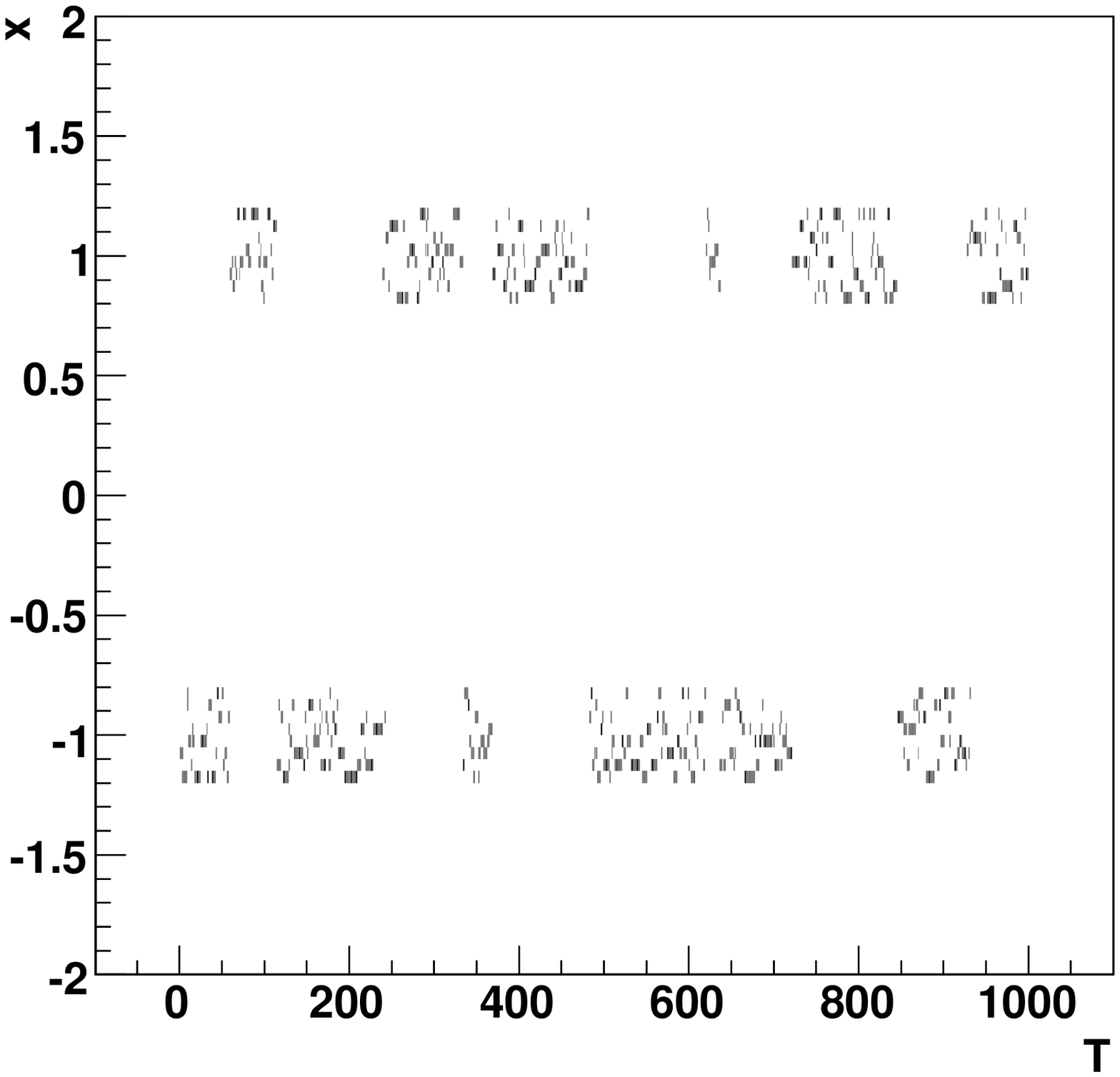}
{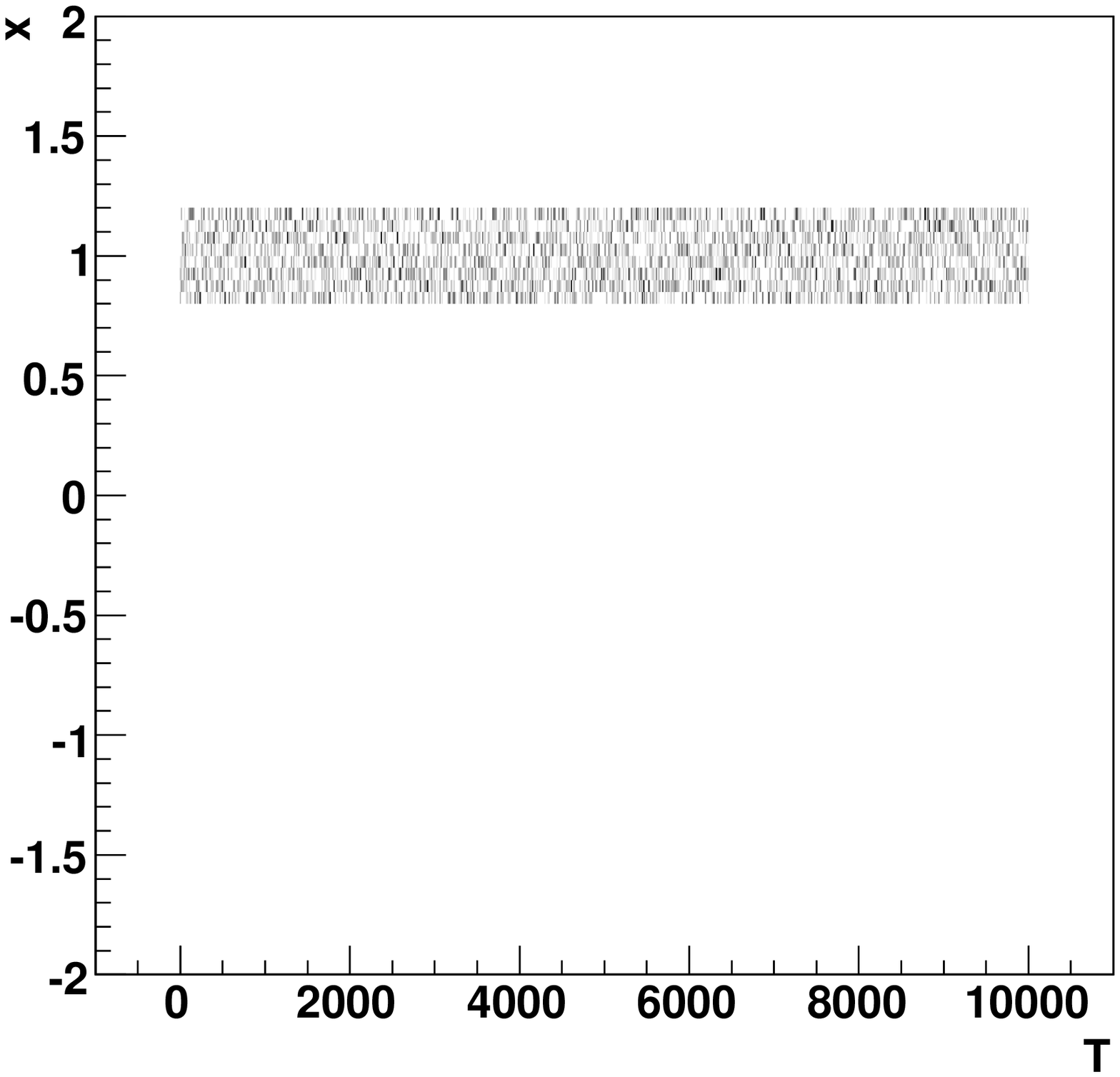}
{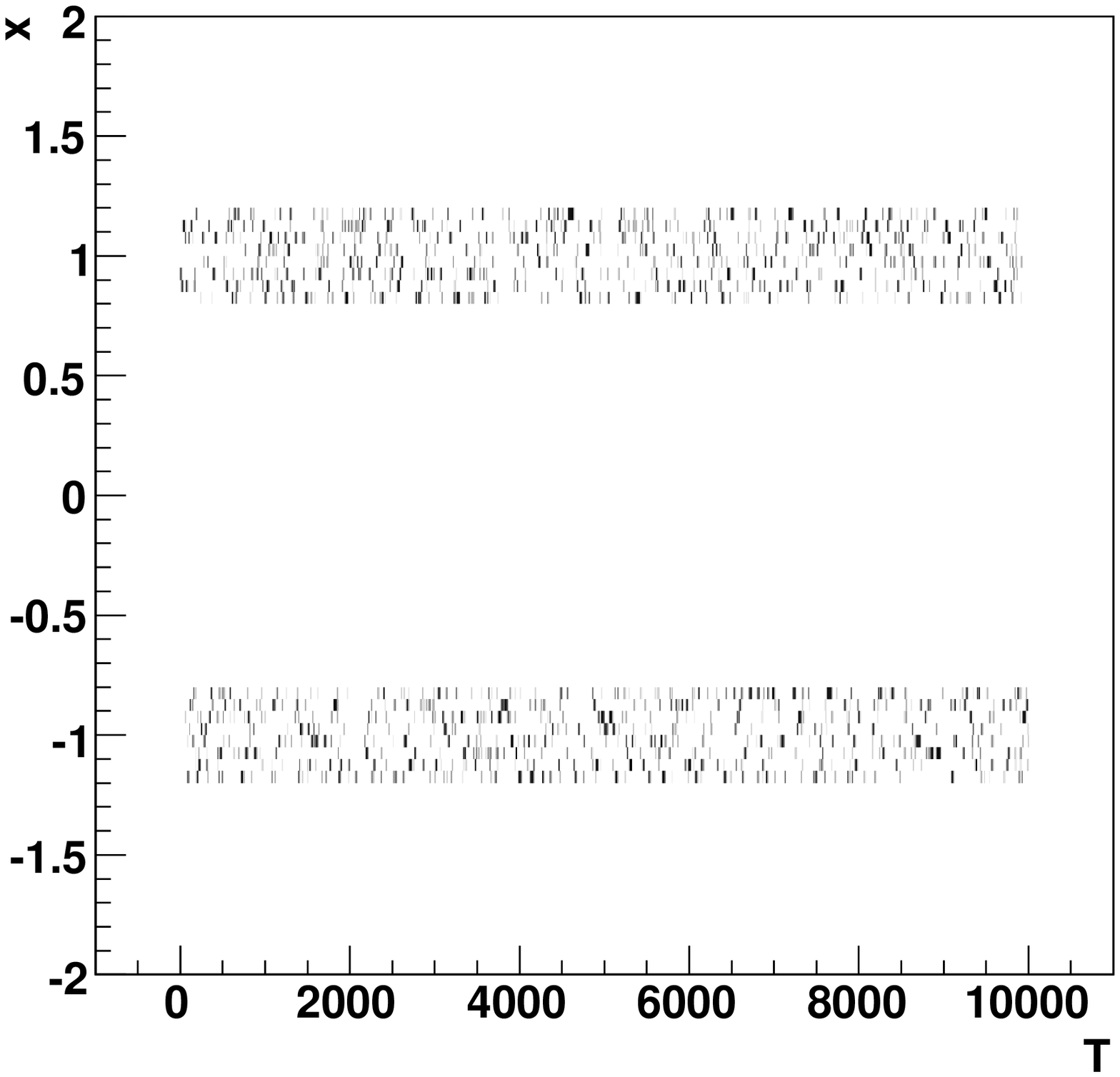}
{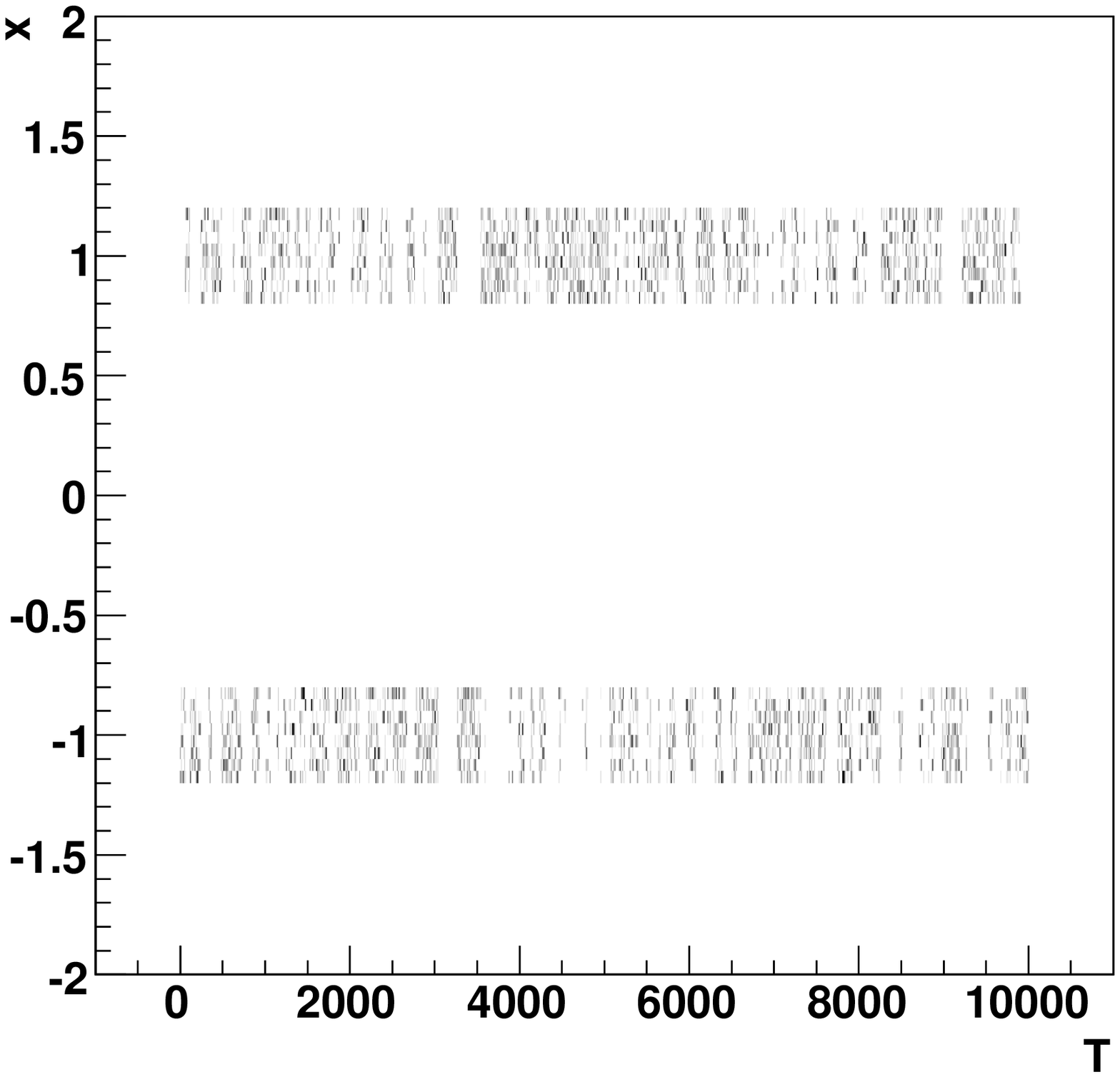}
{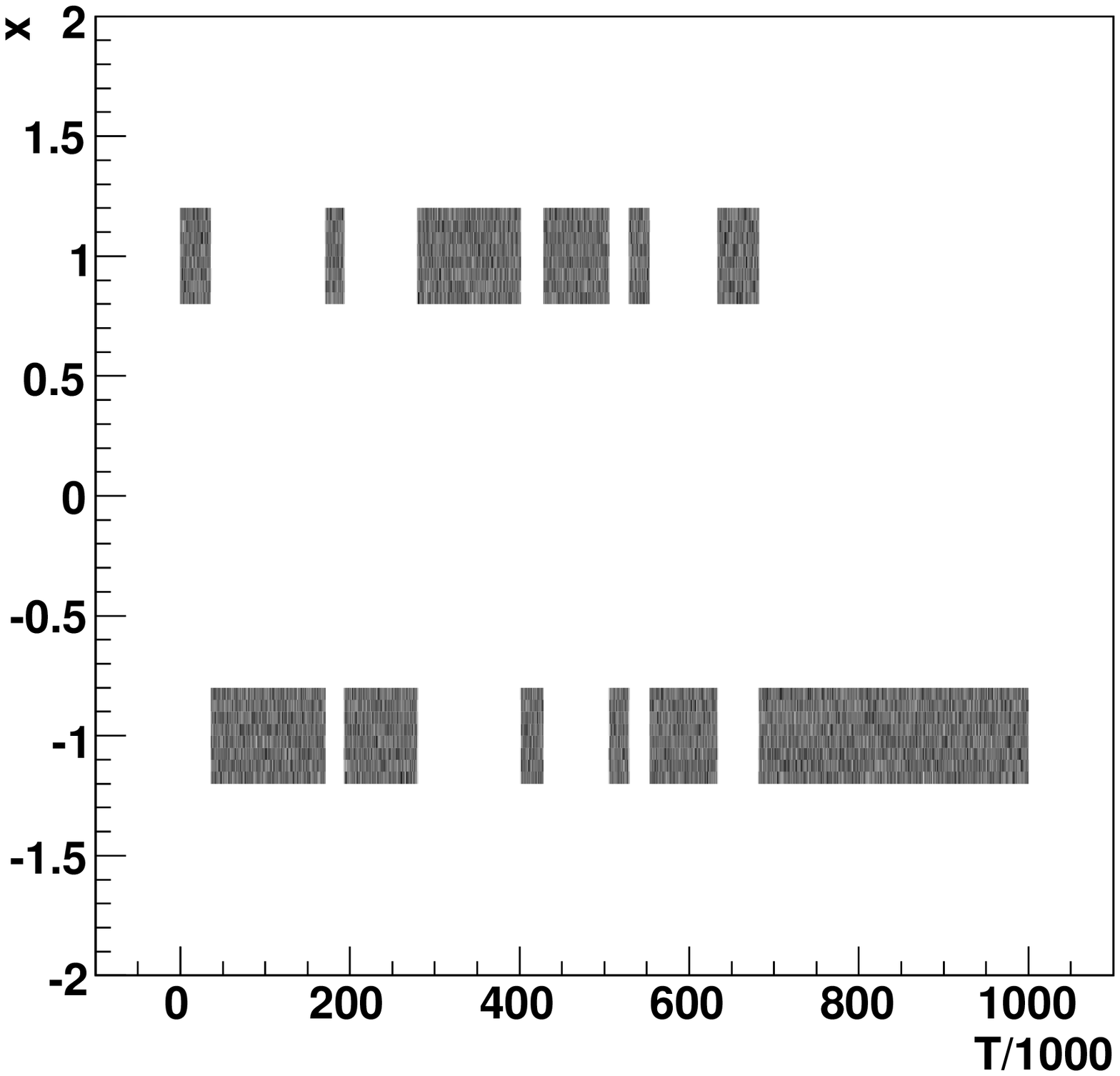}
{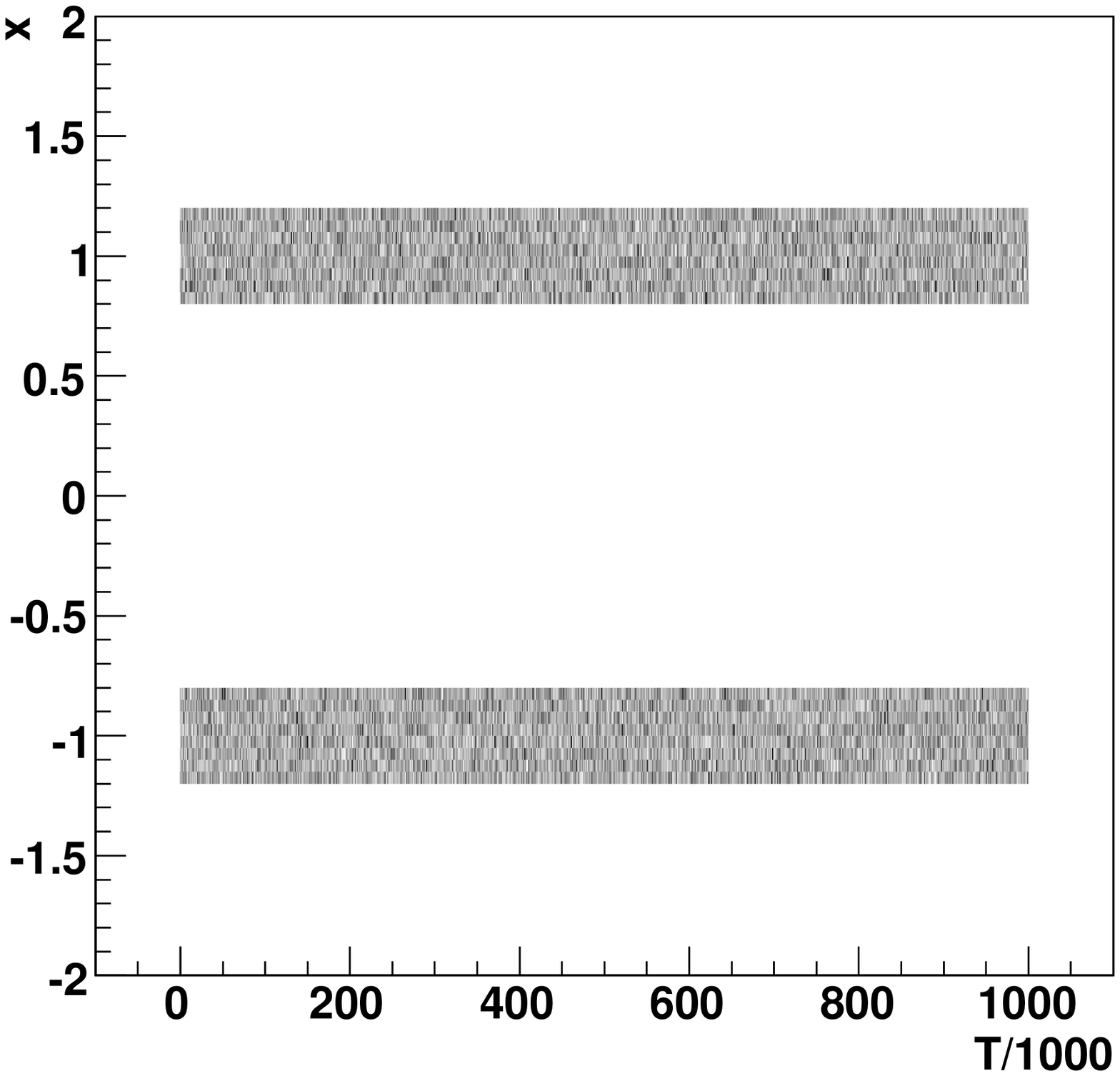}
{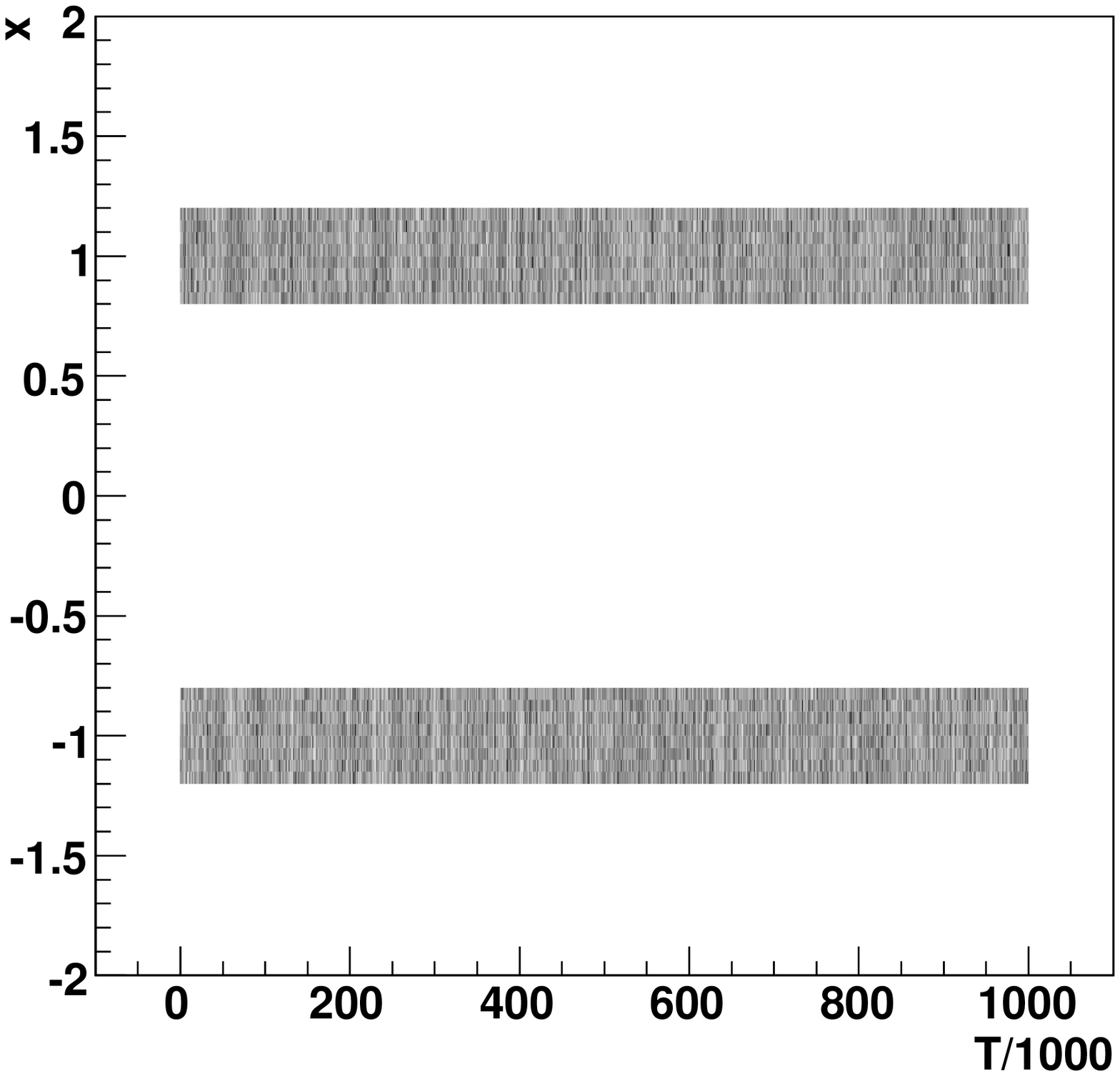}
\caption{\label{fig:1D-time} This figure compares time-series data for
samples from two simple 1D Metropolis samplers (``normal'' in the left
hand column, ``broad'' in the middle column) to data from a bank
sampler (right hand column).  Details of the samplers are given in the
text.  The number of samples increases in each successive row of the
table: one-thousand samples in the top row, ten-thousand samples in
the second row, and one-million samples in the bottom row.  The index
of the sample (i.e.~the ``time'' at which it was generated) is
displayed on the abscissa, while the position of the sample is shown
on the ordinate. Note the rescaling of the time co-ordinate in the
bottom row.}
\end{figure}

\clearpage

This is not to meant to suggest, however, that the ``normal''
Metropolis algorithm is the best algorithm that could be used for the
toy 2D example presented here.  Neither is it the intention to make
the simple Metropolis method look intrinsically bad.  For example the
``broad'' Metropolis algorithm (Figures~\ref{fig:2D-circs}(b), (e) and
(h)), having a length scale of the order of the separation of the two
circle centres, fares much better than the ``normal'' Metropolis
algorithm.  It moves between the two rings well, at the cost of
rejecting a much larger number of proposals between each succesful
step as may be seen from the ``graininess'' in
Figure~\ref{fig:2D-circs}(e).   Quantitatively, the efficiencies with
which the ``normal Metropolis'', the ``broad Metropolis'' and the
``Bank'' samplers made {\em and accepted} new proposals were
respectively $71\%$, $2\%$ and $66\%$. There are plenty of other tricks
that could be played with this particular example using ordinary
Metropolis methods, such as mixing together a narrow Gaussian
distribution suited to exploring the ring ``thickness'' with a broad
Gaussian to allow hopping between rings.  That there are other ways of
making ordinary Metropolis methods work better, however, is not the
point of this comparison.  Such tuning/tinkering will always be valid
and will always lead to better samplers for individual problems.  The
important point to remeber is that the bank sampling algorithm is
designed to remove the need to spend time thinking about part of this
optimisation.  The samples in the ``bank'' of clues already represent
our additional knowledge of the space, and the bank sampling algorithm
(which is in any case just a particular implementation of a Metropolis
method) provides a simple and convenient way of incorporating that
knowledge into a suitable proposal function, thereby promoting the
desired mix of small- and large-scale mobility.

\subsection{Badly skewed clues}

The samples in the bank of clues used by the bank sampler will not
typically be distributed uniformly among the regions of interest or
over the posterior.  In realistic
situations, it could easily be the case that an important part of the
posterior may only feature a small number of times in the ``bank'',
while relatively unimportant regions may be over-represented in the bank.

Though it is not necessary for us to {\em prove} that the limiting
distribution of samples from the bank sampler is not biased by any initial
``skewness'' in the bank'' (this being guaranteed by the $Q$-factors of
the \MHA), it is nevertheless instructive to illustrate this with a
toy example.  We therefore repeat the two-dimensional example of the
previous section but with two changes designed to skew the bank: (1)
an additional ``large'' circle has been added (centred on $(0, 5)$
with radius $3$), and (2) the small circle has been given 10 clues,
the medium-sized circle has been given 5 clues, and the largest circle
has been given only 1 clue.  The results of sampling from this skewed
bank are presented in figure~\ref{fig:2D-circs-THREE-CIRCS}, along
with samplings for comparison made by the and ``normal'' and ``broad''
Metropolis samplers of section~\ref{sec:two-d-toy}.  As expected, the
figure demonstrate the $Q$-factors of the \MHA\ ensure that all three
rings are visited in correct proportion, despite the skewed
distribution of clues. Quantitatively, in this example the
efficiencies with which the ``normal Metropolis'', the ``broad
Metropolis'' and the ``Bank'' samplers made {\em and accepted} new
proposals were respectively $71\%$, $4\%$ and $64\%$.

\subsection{One-dimensional toy problem}

We look at the issue of equilibration between disconnected modes in
more detail in a simpler one-dimensional example in which we choose the
target distribution to be the double top-hat function:
\begin{equation}
f_{1D}({\bf x}) = \begin{cases}
1 & \text{if $|x-c_1|<w_1/2$,}\\
1 & \text{if $|x-c_2|<w_2/2$,}  \\
0 & \text{otherwise.}
\end{cases}
\end{equation}
Figure~\ref{fig:1D-time} compares time-series data for samples from
two simple Metropolis samplers (left hand and middle columns) to data
from a bank sampler (right hand column).  In this example we choose
$c_1=+1$, $c_2=-1$ and $w_1=w_2=0.4$ and with the kernel also of width
$0.4$. Ten bank points were placed under each ``top hat''.  As in the
two-dimensional example the non-bank ``normal'' and ``broad'' samplers
only differ from the bank sampler by having $\lambda=0$ in place of
$\lambda=0.1$ and, in the case of the ``broad'' sampler only, by
having the proposal width $w$ scaled up from $0.4$ to $2$ to match the
spacing between the top-hats.  No burn in time as associated with
these samplings as the initial point was always chosen in the typical
set.

We can see from the right-hand column of Figure~\ref{fig:1D-time} that
the bank sampler is able to make use of the bank points as a means of
hopping back and forth between the two disconnected regions.  In
contrast the ``normal'' Metropolis sampler (left hand column) does not
reach the region at negative $x$ until approximately the 40,000th
sample, and takes even longer to hop back again.

As in the two-dimensional example, we believe the comparison between
the ``normal'' and ``bank'' samplings is the best way of illustrating
the benefits that the bank sampling algorithm brings to the real cases
for which the algorithm was designed.  But again, the comparison is
{\em not} intended to suggest that the ``normal'' Metropolis
implementation is the best single-kernel implementation available for
this example, or that simple Metropolis is intrinsically bad.  On the
contrary, the samplings from the ``broad'' Metropolis implementation
(central column of Figure~\ref{fig:1D-time}) are at least as good as
those from the bank sampler.  The benefit of the bank sampler,
however, is again that it aims to provide a large amount of this
proposal-function-optimisation automatically.  Quantitatively, the
efficiencies with which the ``normal Metropolis'', the ``broad
Metropolis'' and the ``Bank'' samplers made {\em and accepted} new
proposals here were respectively $40\%$, $13\%$ and $37\%$.

\begin{figure}
\threegraphst
{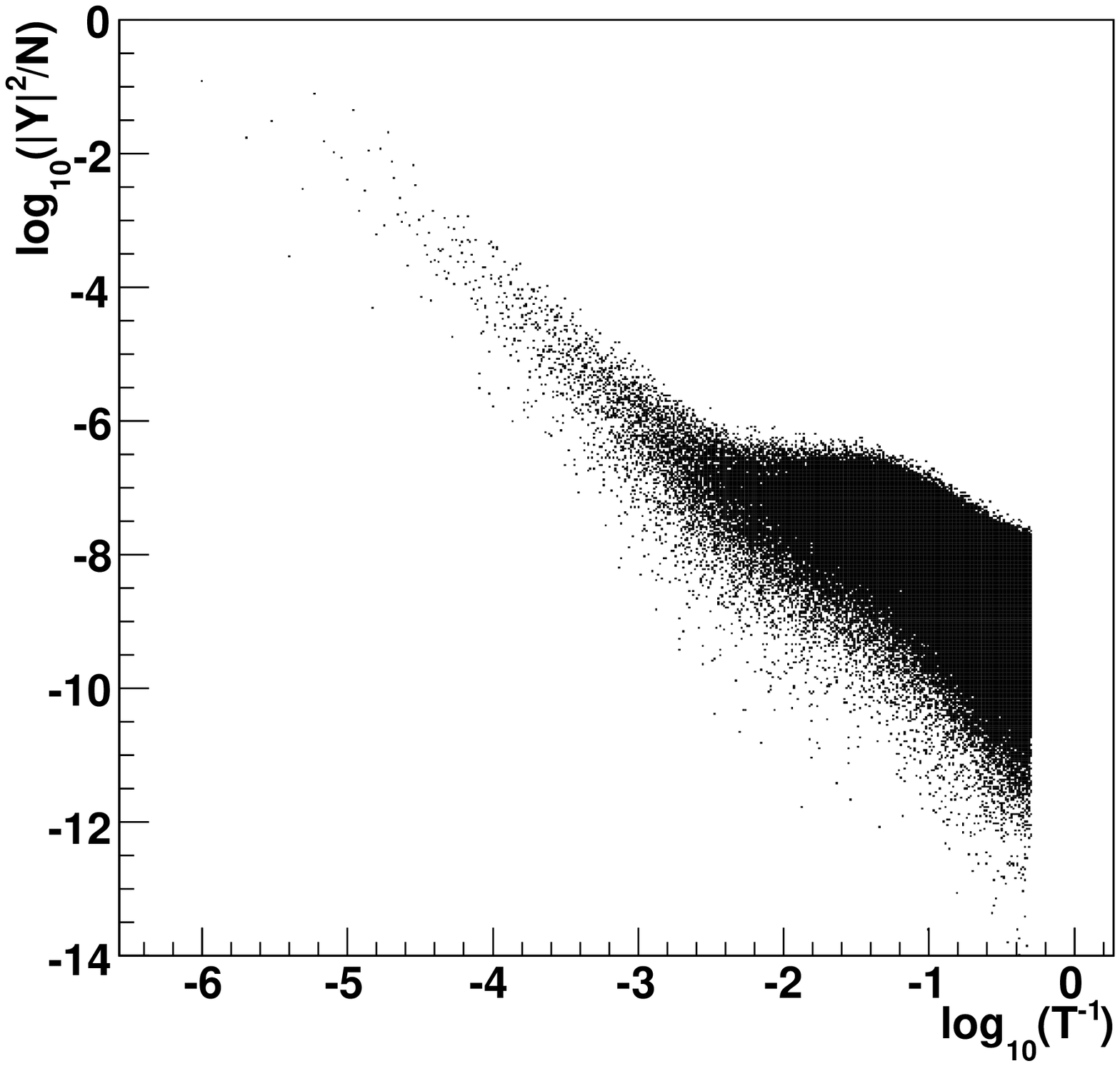}
{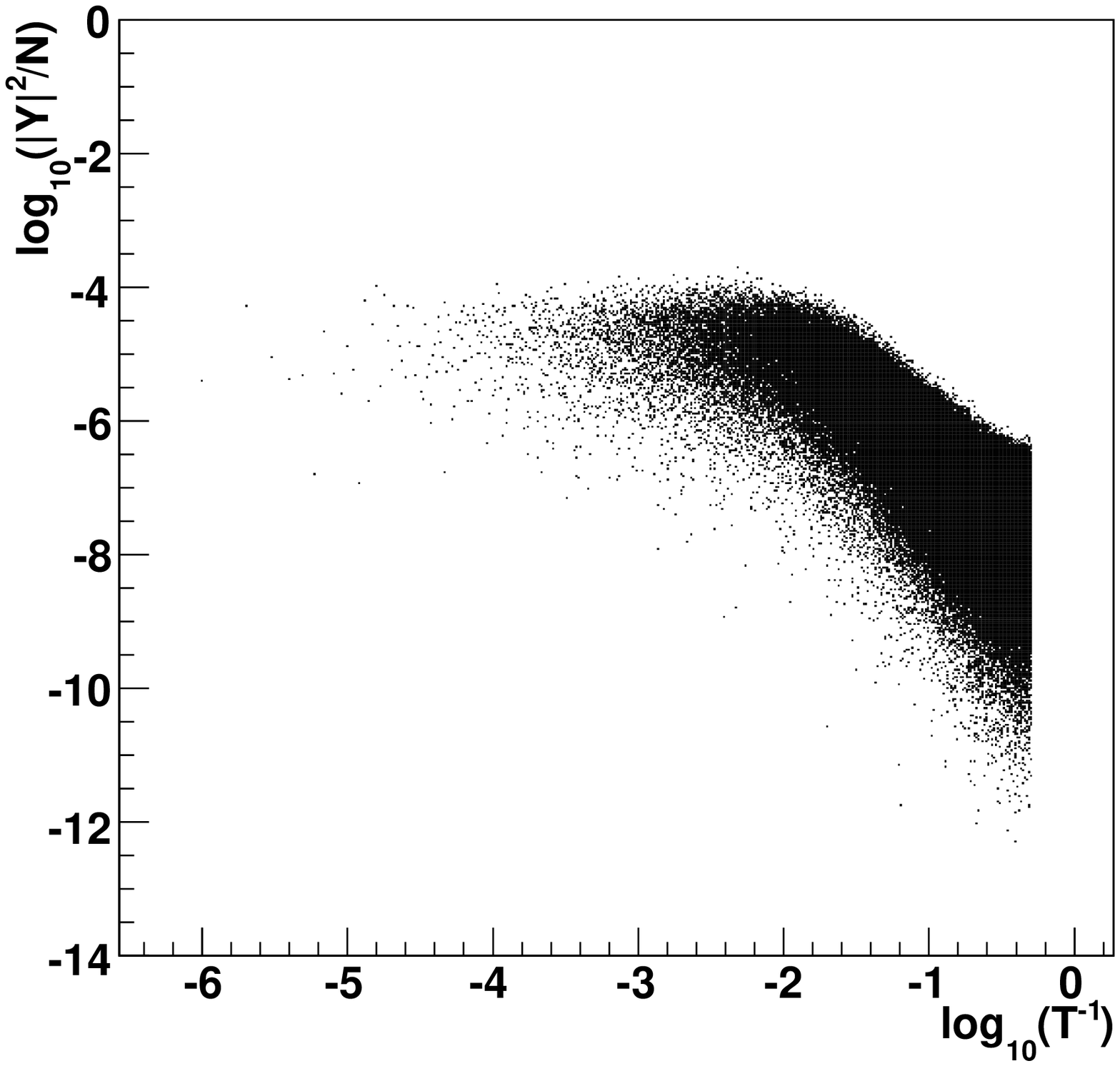}
{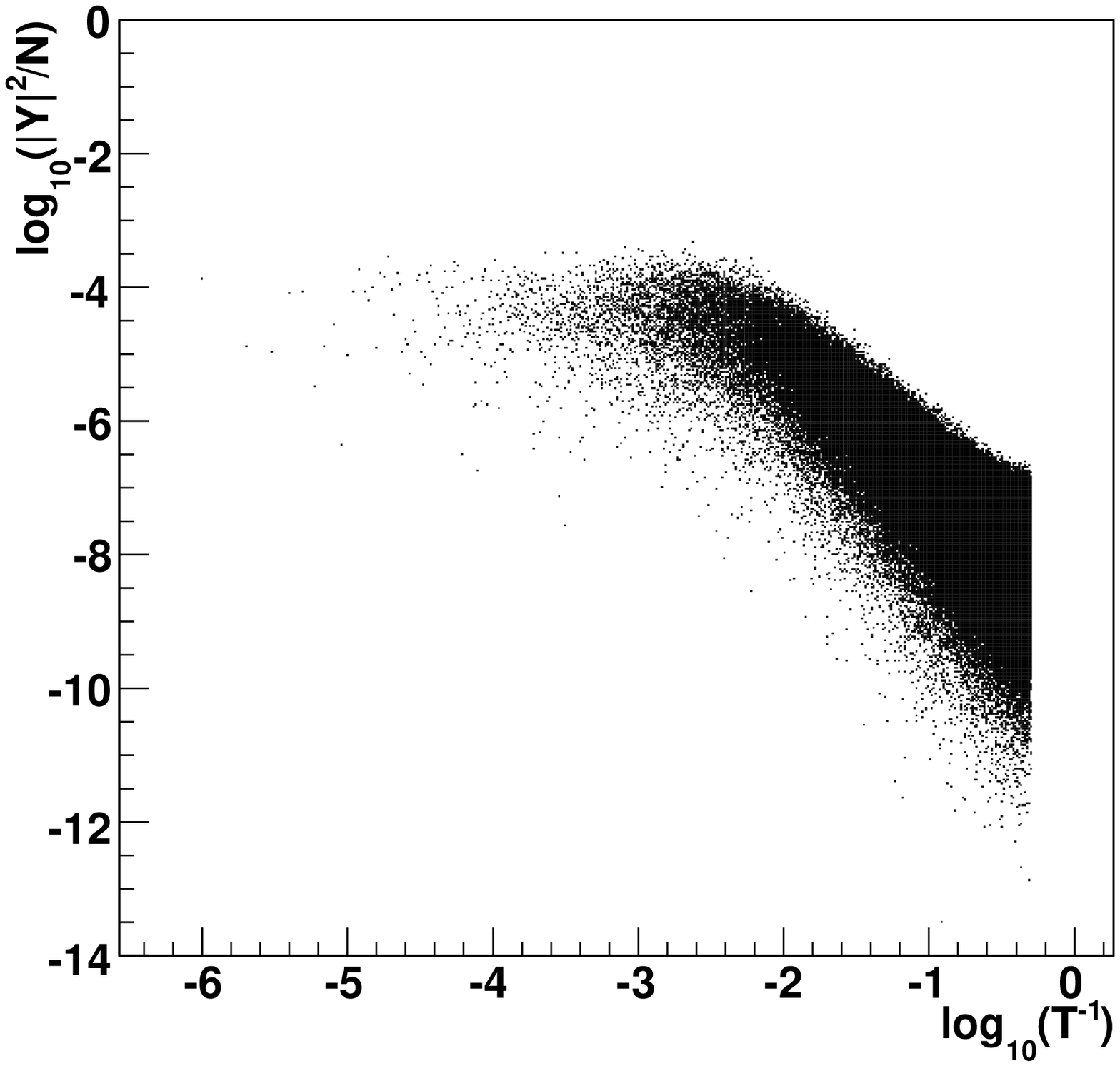}
\caption{Histograms for the one-dimensional examples described in the
  text, indicating the timescales on which samples can be considered
  independent.  Appendix~\ref{app:convergenceDef} describes how these
  histograms are defined and how to interpret them.\label{fig:1D-conv}
  }
\end{figure}

We can compare the time-scale over which samples from the various
samplers become statistically independent using ingredients described
in appendix~\ref{app:convergenceDef} from the method of
\cite{Dunkley:2004sv}.  The authors would like to point out
that the tools provided in \cite{Dunkley:2004sv} for assessing
convergence go far beyond those ingredients which are used here.  The
time-scale over which samples become statistically independent is only
one {\em part} of those tools.  The exact set of tests one would want
to use in a real situation to assess convergence/mixing or typicality
of samples would depend on their eventual intended use.  For example,
if the samples were to be averaged to estimate a mean, then one {\em
could} identify a set of independent samples by determining the
time-to-independence and could then thin the samples and take an
expectation using the thinned set.  As \cite{Dunkley:2004sv} points
out, however, a much better strategy is not to thin the samples, but
to use all the samples, uncorrelated or otherwise, provided that
correlated samples can be shown not to bias the chain ouput. Assessing
when this can be done requires {\em both} that a flat region exists in
the power spectrum implying a clear measurement of the de-correlation
time, {\em and} that the ratio of power at large scales to the number
of samples is sufficiently small.  (See section 4.3 of
\cite{Dunkley:2004sv}).  We do not ourselves use our toy-chains for
any purposes other than (1) measuring efficiencies, (2) projecting the
chains for visual inspection, and (3) looking at biases from chains
which have not yet run for long enough (in Figure
~\ref{fig:muconstraint1}).  In this section we do not, therefore,
invoke the full power of the tests available in \cite{Dunkley:2004sv}.
We investigate only the power spectrum and the de-correlation time.

Figure~\ref{fig:1D-conv}(c) shows that the bank sampler underwent
correlated random-walk behaviour on timescales of $10^2$ samples and
shorter, but generated independent samples at timescales of $10^3$
samples and above where the power spectrum is flat.  In contrast,
Figure~\ref{fig:1D-conv}(a) for the ``normal'' Metropolis sampler
shows no clear flat region at long timescales -- even after $10^6$
samples.  This decorrelation measure is reporting that by this time
the ``normal'' 1-D Metropolis algorithm is, at best, {\em only just}
beginning to take samples from the two regions in an independent
manner, in agreement with Figure~\ref{fig:1D-time}(g).
Figure~\ref{fig:1D-conv}(b) clearly shows that the ``broad'' 1-D
Metropolis algorithm can produce uncorrelated samples on the same
timescale as the bank sampler, and would be much better than the
``normal'' one-kernel Metropolis algorithm if one were seriously
trying to sample from this example in that way.

The important result here is not that the ``normal'' 1-D Metropolis
algorithm is only just beginning to produce samples representative of
the whole distribution (this result is to be expected as the chosen
step size is there, by design, too small to allow easy jumps between
the two regions!) but (1) to demonstrate the time scale on which the
bank sampler's samples become effectively independent and
representative of the distribution as a whole, and (2) to demonstrate
a case in which the test identifies samples which have not yet lost
their correlated random walk behaviour.\footnote{Of course this test can only
make ``local'' statements about the samples. If there were an
important region which had never been visited by the chain, this test
would be unable to point out that all existing samples are correleted
by their non-membership of this region.}  It should be obvious that
one cannot use exclusively narrow proposal distributions on examples
like this, any more than one can expect them to work on the similar
but more realistic distributions for which the bank sampler was
created.  Intelligent tuning, either by hand, or by bank sampling or
by some other means, will always be required.

\subsection{Safety with respect to free parameters}

\begin{figure}
\epsfig{file=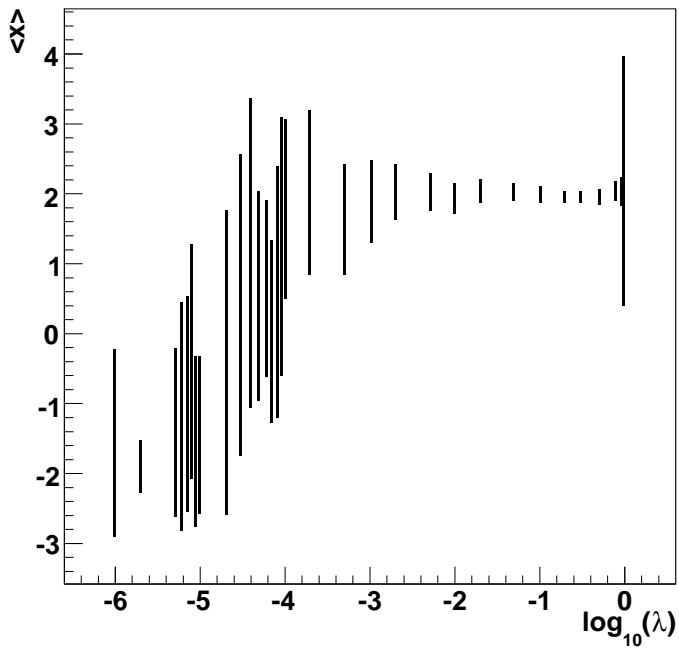, width=4in}
\caption{Ensemble measurements of the average value of the
  $x$-coordinate for the two-dimensional example are plotted for
  various values of ``$\lambda$''.  Note that $\lambda$ represents the
  proportion of ``bank'' (as opposed to simple Metropolis) proposals.
  Every step toward the left of the plot therefore represents an order
  of magnitude reduction in the probability of making a ``bank''
  proposal, thus tending to the limit of ordinary Metropolis sampling.
  For each value of $\lambda$, ten independent Bank samplers were set
  running, each with ten bank points randomly distributed around each
  circle (making 20 bank points in total).  Each sampler generated
  200,000 samples.  From the ensemble of measurements of
  $\left<x\right>$ thereby obtained, a mean value of $\left<x\right>$
  and a root-mean-square deviation of these measurements from their
  mean were obtained and plotted as the vertical error bars in the
  figure indicating ``mean''$\pm$``RMS''.  The true value of
  $\left<x\right>$ is $+2$ for the two-dimensional example.  Though
  each sampler was independent, the starting value for each sampler
  was a random point on the circle at negative $x$ centred on $x=-2$.
  The failure of ``ordinary'' Metropolis to escape its initial
  conditions in the time available is evident at values of $\lambda$
  below ${10}^{-3}$.  In this example, values of $\lambda$ above
  ${10}^{-3}$ are seen to work well.
  \label{fig:muconstraint1}}
\end{figure}

It would be reassuring to demonstrate that free parameters
like $\lambda$ do not need to be tuned excessively in order to make
the bank sampling algorithm useful.  In the earlier examples we took
$\lambda$ to have the value $0.1$, having said that it {\em must} lie
somewhere in $0<\lambda<1$.  How dependent is the sampling quality on
$\lambda$ ?  We investigate the two extremes $\lambda\rightarrow 0$
and $\lambda \rightarrow 1$ separately.

Figure~\ref{fig:muconstraint1} shows, in the context of the
two-dimensional example, how close to $0$ the value of $\lambda$ can
become before the ugly behaviour ({\em i.e.}~poor converage 
properties) of standard Metropolis sampling are revealed in runs
constrained to 200,000 samples.  Figure~\ref{fig:muconstraint1}
demonstrates that values of $\lambda$ can be taken as low as ${10}^{-3}$
(only one bank proposal in every $10^3$ proposals) and still provide
vast improvement over standard methods.  This result is consistent
with the rule-of-thumb that of order 100 ``bank'' proposals must be
accepted if one is to see any benefit from the bank sampling
technique, thereby making it pointless to try values of $\lambda$ below
$100/N$ where $N$ is the number of samples.  Where bank points are
atypical of the target distribution, the practical lower bound for
useful $\lambda$ will be higher.
 
\begin{figure}
\epsfig{file=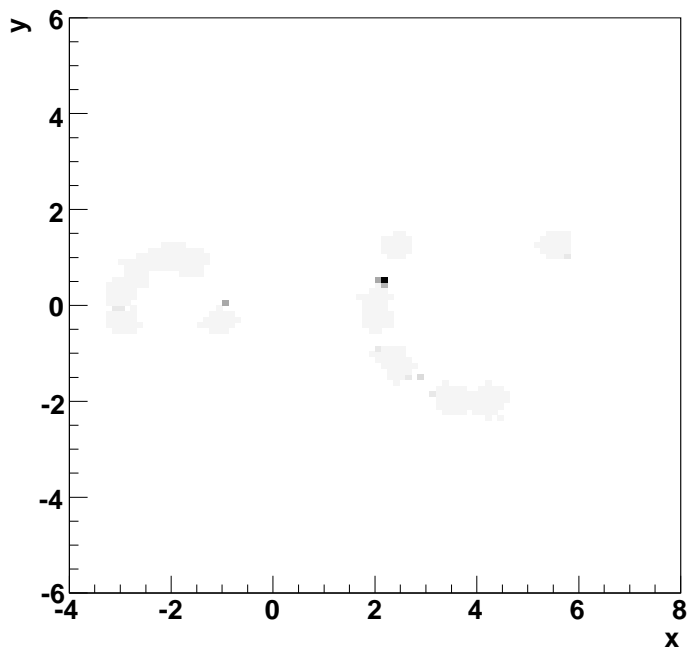, width=4in}
\caption{This is a sampling of 10,000 points from the two-dimensional
  example using a bank sampler with $\lambda=0.999$ and with ten bank
  points distributed randomly around each circle. The locations of the
  bank points can be inferred from the faint clouds of low-weight
  points they have attracted.  It is evident, however, that in the
  10,000 points seen so far, almost all the weight of the sampling so
  far has been concentrated in the single point located near
  $(2.2,0.5)$.\label{fig:badlowmu}}
\end{figure}

It is tempting to assume that figure~\ref{fig:muconstraint1} gives a
 green light to arbitrarily large values of $\lambda$ also, but this
 would be a mistake.  The apparent agreement with the correct value of
 $\left<x\right>=2$ at values of $\lambda$ close to $1$ is a
 misleading feature of that plot.  In the $\lambda\rightarrow 1$
 limit, the bank sampling proposal distribution becomes {\em entirely
 non-local} thus making the sample very close to importance sampling.
 Inevitably, Gaussian kernels centred on the ten bank points around
 each circle in the two-dimensional example form a very bad
 approximation to the true form of equation~\ref{eq:twodfunc}, and
 some important parts of the target distribution are many standard
 deviations away from the centre of the nearest kernel.  In this
 limit, therefore, the bank sampler should be expected to spend a
 large fraction of its time stuck repeatedly re-sampling points in
 such regions, making it highly likely that the entire sampling will
 be dominated by a single point.  We can see this effect in
 figure~\ref{fig:badlowmu}.  It is clear to the eye that this sampler
 has not yet done a good job by any standard.  There is a large part
 of the bulk of the target distribution near $(6, -1)$ which is
 approximately 20 standard deviations away from the nearest bank point
 (the kernel width continuing to be 0.1) and so we can estimate that
 in the limit $\lambda\rightarrow 1$ it will take of the order of
 $\exp(20^2/2)\sim10^{87}$ samples to be drawn before this region is
 visited.  This is a classic example of why importance sampling fails
 in large numbers of dimensions, and serves to make it clear why the
 \MHA\ is so commonly used in its place.  At this value of $\lambda$
 almost no ``standard'' local Metropolis steps are being proposed, and
 so there is no quick way for the sampler to wander into the regions
 that are not in the vicinity of bank points.  The only way the
 sampler can compensate for this lack of mobility is to sit for long
 periods of time at the places it knows it will be unable to reach
 again, to ensure that the right distribution is obtained as
 $t\rightarrow\infty$.

\begin{figure}
\epsfig{file=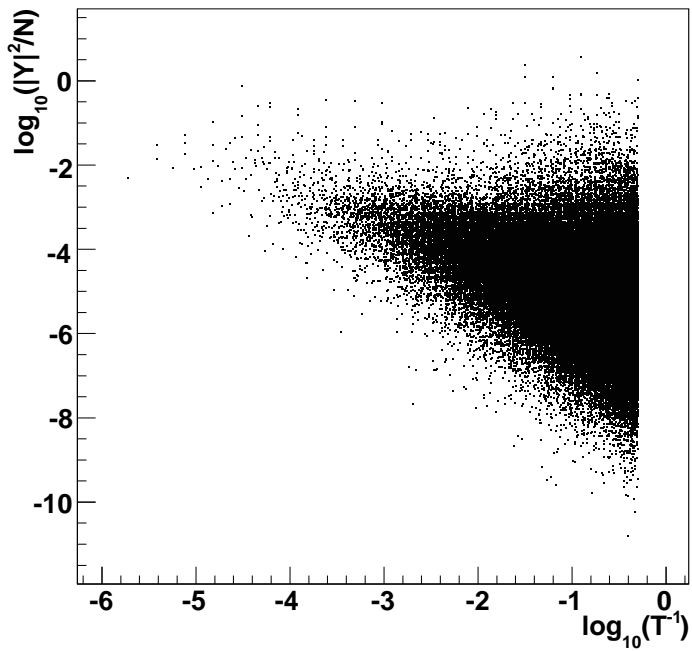, width=4in}
\caption{Here we show the power spectrum of the $x$-components of the
  samples of figure~\ref{fig:badlowmu} (extended to one million
  samples) using the same technique as in figure~\ref{fig:1D-conv}.
  The power spectrum is consistent with slope $-2$ and with no
  evidence for a flat region at large timescales.  There is therefore
  no evidence of loss of random walk behaviour and no evidence of a
  transit into a region in which samples become statistically
  independent. \label{fig:convoflowmu}}
\end{figure}

Fortunately, we already know that a finite admixture of simple
Metropolis steps will avoid the importance sampling pitfall, making
the algorithm at worst only $(1-\lambda)$ times as efficient as
standard Metropolis.  The upper bound on sensible choices of $\lambda$
is then a function not of the number of samples $N$ but a function of
how inefficient you are prepared to make the sampler in the event that
the bank points are not very good.  It is unlikely therefore that
values of $\lambda$ above $0.9$ are likely to be of any use -- unless
the banked points were already so good that standard Metropolis steps
(slowed down by being only used 10\% of the time) were not required to
traverse poorly represented parts of the target distribution.  Even if
one were foolish enough to avoid this advice and proceed with a
dangerously high value of $\lambda$, it seems likely that the
resultant poor performance of the sampler would be spotted by standard
techniques.  For example figure~\ref{fig:convoflowmu} shows the
Fourier power spectrum of the $x$-coordinate values of the example of
figure~\ref{fig:badlowmu}, after extension to a total of one million
samples.  It is clear that there is no evidence for a horizontal
component to the spectrum at the top left, and that the whole spectrum
is indicative of slow random walk (in the vicinity of the high-weight
point).  This sample would not be considered acceptable by this
measure.\footnote{Of course, no test can successfully identify cases
in which a method had failed to discover one or more isolated region,
but has covered a subset of the desired regions well.}

To summarise, it is expected that in most reasonable problems there
will be a large dynamic range of sensible values that $\lambda$ can
take, and the range can be estimated from consideration of the desired
length of chain and the worst case inefficiency one is prepared to
accept.  In the two-dimensional example, values of
$\lambda\in[0.001,0.9]$ would have been fine.


\subsection{Multi-dimensional examples \label{sec:mde}}

Using the performance of any algorithm on a multi-dimensional example
as a measure of its likely suitability for use on other
multi-dimensional problems can be very unreliable.  Any number of
small differences between the real and an example target distribution
or the real and an example proposal function can lead, in high numbers
of dimensions, to large positive or negative changes in an algorithm's
suitability.  Rather than introduce a toy multi-dimensional example of
dubious generality, we instead refer the reader to the paper
\cite{lester-allanach-prior} for which the bank sampler was developed.
The analysis in that paper required a sampler that could sample from a
parameter space of eight real dimensions and one binary parameter,
sign$(\mu)$.  There were two completely disconnected regions of
interest (corresponding to sign$(\mu)=\pm 1$) in addition to the
presence of two modes within the sign$(\mu)=-1$ space.  The bank
points were generated by running twenty independent chains.  The
initial 2,000 ``burn-in'' points of each of these chains was
discarded.  Ten chains, for sign$(\mu)=1$, were 40,000
step\footnote{When we discuss ``number of steps'', we refer to the
number of attempted points in a chain. This is equal to the number of
function evaluations, and therefore proportional to the CPU time
necessary for the sampling.} standard Metropolis samplings from the
target distribution each starting from random widely separated
points. The other ten chains, for sign$(\mu)=-1$ had 20,000 steps,
also statistically independent.  A sub-sample of these points (5,000
different points, picked at random from the ten burnt-in chains) was
used to form the bank. We define efficiency by the number of {\em
disjoint} points sampled divided by total number of attempted points.
The efficiencies in the initial stage of determining the bank points
and at the final bank sampling stage, were approximately identical
(e.g.\ 6$\%$ for flat priors) because we had allowed the initial-stage
chains to burn-in and so the bank points had reasonably large
posterior densities.  We may estimate the total additional overhead
due to bank sampling, including the initial runs and associated
burn-in period, to be 640,000 / 5,000,000 = {\bf 12.8}$\%$, since the
efficiencies are approximately equal in the initial and final-stage
chains. This was deemed to be a small overhead on the usual sampling.
Therefore, the additional overhead was not optimised in any way: it is
very possible that a different strategy would yield a smaller
overhead, for example by running shorter chains in the initial stages.
\begin{figure}
\epsfig{file=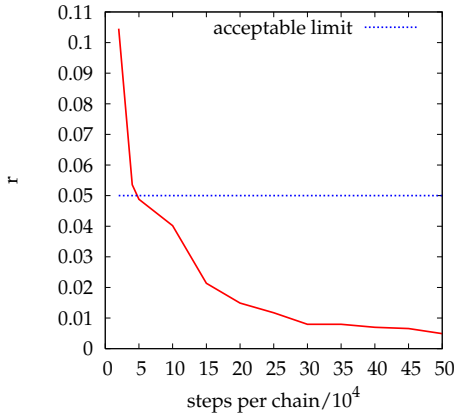, angle=270, width=0.8 \wth}
\caption{Mixing as measured by $r$ as a function of number of steps
  in each bank chain for the CMSSM fits with flat priors in
  ref.~\cite{lester-allanach-prior}. Note that ``steps per chain'' refers to
  the number of {\em attempted} steps.
  \label{fig:rhat}}
\end{figure}
We estimate that the separation of modes would mean roughly an additional
order of magnitude overhead with a ``standard'' sampler.
The small overhead of 12.8$\%$ for bank sampling should therefore be (very
favourably) compared with this estimate.
We may use the Gelman-Rubin estimate $\hat R$~\cite{stat}
to investigate mixing derived from statistically
independent chains. $\sqrt{{\hat R}}-1$ provides an estimated upper bound on
the
fractional decrease
in root variance of any scalar quantity that could be obtained by running the
MCMCs
for more steps. Here, we calculate the maximum such estimate from all of the
eight input parameters in the SUSY run, $\rhatmax$. We
take values of
$r \equiv \sqrt{\rhatmax}-1$ less
than 0.05 to indicate adequate mixing. Fig.~\ref{fig:rhat} demonstrates that
adequate mixing was already obtained by around 50,000 steps per banked chain.
We did not generate data for the standard non-banked approach in
ref.~\cite{lester-allanach-prior} except in the
initial stages, where (in 20,000 steps) $r=2.16,0.52$ for
sign$(\mu)=-1,+1$ respectively. For comparitive
purposes, for the same number of steps, the banked method yields
$\rhatmax - 1=0.11$, an order of magnitude improvement in the mixing. Thus, at
least in the early running of the MCMC, the banked approach yields much better
mixing.
In order to compare mixing behaviour in the later evolution of the MCMC, we
look to previous
similar fits that were performed for sign$(\mu)=1$, Fig.~1 of
ref.~\cite{Allanach:2005kz}. Values of $r<0.05$ were reached for 600,000 steps,
an order of magnitude more than was obtained via the banked method (which only
has a 12.8$\%$ overhead in CPU time).

\section{Conclusions}

A method for sampling from multi-modal distributions in moderate
numbers of dimensions has been presented.  To be of benefit over
standard algorithms the sampler must recycle ``old'' data from earlier
anlayses or aborted samplings, or make use of educated guesses about
the nature of the space to be sampled.  With such a bank of ``clues''
the algorithm stands a good chance of (1) equilibrating between
isolated modes much faster than standard algorithms, and (2) taking
less time to generate uncorrelated samples.  When circumstances suit
the sampler it may be one or more orders of magnitude faster than
competitors.  If the bank of clues happens to be no good, the
performance penalty of using the sampler is tunable, bounded, and is
typically no more than 10\%.  The implementation overhead is confined
to the sampling-from and evaluation-of a proposal function only mildly
more complex than that used in the simplest forms of the
Metropolis-Hastings algorithm -- it being a weighted sum of such
proposals. No advanced operations (eg matrix manipulations, Fourier
transforms) are required, making it one of the simplest non-standard
sampling algorithms to implement.  The algorithm has been shown not to
require heavy tuning of the most important free parameter $\lambda$ --
the proportion of bank proposals. 

\par

The bank sampling method could be tried with other samplers that are
not of the Metropolis-Hastings type, for example the popular
Hamiltonian Monte Carlo algorithm~\cite{ham}. The only change to our
algorithm would be trivial: $(1-\lambda)$ would become the probability
of a Hamiltonian Monte Carlo step not of a Metropolis-Hastings one.

\par

A different algorithm designed to address the issue of sampling from
multi-modal distributions has recently been proposed proposed within
the astrophysics community \cite{Feroz:2007kg}. This algorithm and our
own are both well suited to parameter estimation.  In order to
function efficiently, however, the algorithm of \cite{Feroz:2007kg}
must enclose certain parts of the posterior probability mass within
one or more bounding ellipsoids, containing little ``empty space''.
In cases where the bulk of the probability lies within thin sheets or
hypersurfaces within a larger space (as was the case in our
multi-dimensional example \cite{lester-allanach-prior} and is typical
for particle physics constraints involving recent cosmic microwave
background data) it seems likely that achieving this coverage would
either require (1) a small number of large ellipsoids, leading to very
low efficiency at the sampling stage, or (2) an unfeasably large
number of tiny ellipsoids to allow a sufficiently close representation
of the surface.  In contrast, the Bank Sampler needs only to ``seed''
such a surface with a small number of bank points (whose kernels need
not overlap) before allowing the ``ordinary'' Metropolis steps to fill
in between the gaps.  It seems likely, therefore, that the ``bank''
sampler may provide a simpler method of attacking surface-like
constraints.  It is not immediately clear which method would fare
better for isolated Gaussian modes, but the method of
\cite{Feroz:2007kg} is certainly well optimised for that regime.
Finally we note that \cite{Feroz:2007kg} provides a feature not
present in the ``bank'' sampler (or indeed in any other \MHA) which is
a fast calculation of the {\em Bayesian evidence}\footnote{See
\cite{MacKay} or any other Bayesian textbook} as a by-product.

\section{Acknowledgements}

This work has been partially supported by the United Kingdon's Science
and Technology Facilities Council.  Thanks is given to D.J.MacKay
(Cambridge Inference Group) and members of the Cambridge Supersymmetry
Working Group for useful discussions.

\appendix

\section{Measuring the number of steps
      between independent samples in the chain.}

\label{app:convergenceDef} 

Following the method of \cite{Dunkley:2004sv} we assess time (that is
to say the ``number of steps'') between which samples in a Markov Chain
become independent by
looking at the Fourier power spectrum of the path taken by each
sampler.  Given the full set of $N$ samples $\left\{x^{(0)}, x^{(1)}, ... ,
x^{(N-1)} \right\}$ from the chain, we define the discrete Fourier
transform:
\begin{equation}
Y(T^{-1}) = {\frac 1{\sqrt N}} \sum_{n=0}^{N-1} x^{(n)} \exp \left[ -i 2 \pi n (T^{-1})\right]
\end{equation}
in terms of a discrete inverse timescale variable $T^{-1}$ which may
take values in $\left\{0/N, 1/N, 2/N, ..., (N-1)/N\right\}$.  The
strength of correlation between points which are $T$-samples apart
(i.e.~spearated by $T$ units in ``time'') is given by the power
${\left|Y(T^{-1})\right|}^2$.  A sampler generating truly independent
events should have a power spectrum that is flat across all
timescales.  In contrast, a random walk (non-independent samples)
should have a non-flat power spectrum with greater power at long
timescales.  From \cite{Dunkley:2004sv} random walk behaviour should
be evident as a straight-line power spectrum of slope $-2$ on a plot of
$\log_{10}(\left|Y\right|^2/N)$ against $\log_{10}(T^{-1})$.
Independent samples should be evident on the same axes as a flat power
spectrum.  A sampling algorithm, such as the Metropolis Algorithm,
which produces highly dependent samples at small timescales but
eventually produces independent samples at long timescales should have
a power spectrum containing both parts -- a flat region at large times
turning over into the gradient $-2$ part at short times.  The presence
of the two regions can therefore be used to indicate whether
convergence has been achieved (in the sense of samples appearing to
have become independent at the longest timescales of the problem) and
the position of the junction between the two domains indicates the
timescale at which this state was achieved.

\end{document}